\documentclass[12pt]{article}
\usepackage[cp1251]{inputenc}
\usepackage{inputenc}
\usepackage{geometry}
\usepackage{bbm}
\usepackage{amsfonts}
\usepackage{mathrsfs}
\usepackage{latexsym}

\usepackage{epsfig}
\usepackage{epstopdf}
\usepackage{graphicx}
\usepackage{amssymb}
\usepackage{amsmath}
\usepackage{dcolumn}
\usepackage{bm}
\usepackage{color}
\usepackage{comment}
\usepackage{xcolor}
 \usepackage{amssymb,amsmath,epsfig}
\begin{document}
\title{\textbf{Matter Accretion onto a Conformal Gravity Black Hole}}
\author{G. Abbas$^1$ \thanks{ghulamabbas@iub.edu.pk}
and A. Ditta$^1$ \thanks{adsmeerkhan@gmail.com}
\\$^1$Department of mathematics, The Islamia University of Bahawalpur,\\
Bahawalpur-63100, Pakistan.}
\date{}
\maketitle
\begin{abstract}
The accretion of test fluids flowing onto a black hole is investigated.
Particularly, by adopting a dynamical Hamiltonian approach, we are capable to find
the critical points for various cases of black hole in conformal gravity.
In these cases, we have analyzed the general solutions of accretion
employing the isothermal equations of state.
The steady state and spherically symmetric accretion of different test fluids onto the conformal gravity black hole
has been considered.
Further, we have classified these flows in the context of equations of state
and the cases of conformal gravity black hole. The new behavior of polytropic fluid accretion is also discussed in all three
cases of black hole. Black hole mass accretion rate is the most important part of this research
in which we have investigated that the Schwarzschild black hole produce a typical signature than the
conformal gravity black hole and Schwarzschild de-Sitter black hole.
The critical fluid flow and the mass accretion rate have been presented graphically by the impact parameters
$\beta$, $\gamma$, $k$ and these parameters have great significance.
Additionally, the maximum mass rate of accretion fall near the universal and Killing
horizons and minimum rate of accretion occurs in between these regions.
Finally, the results are compared with the different cases of black hole available in the literature.
\end{abstract}
{\bf Keywords:} Accretion; conformal gravity BH; Schwarzschild de-Sitter BH; Schwarzschild BH.\\
{\bf PACS numbers:}04.40.Dg,95.30.Sf,04.50.Gh
\section{Introduction}
Einstein's theory of gravity was suggested at the end of $1915$ and is still
the standard framework for the account of the chrono-geometrical structure and
the gravitational fields of spacetime. Despite its certain success to account
for a comprehensive observational data \cite{1}, the theory is troubled by
some important issues that surely point out the entity of new physics.
One of these issues is the existence of spacetime singularities in the physically
applicable solutions of the Einstein equations. At a singularity, capability of prediction
is lost and the standard physics breaks down. It has been suggested that the
issue of spacetime singularities in the Einstein theory of gravity can be resolved by the familiar theory of quantum gravity. Since, the present understanding of physical laws cannot assume a singularity and there are various efforts to solve the singularity issue
\cite{1e}-\cite{1p}. One of the effort to solve the singularity issue has been proposed in \cite{1m},
where, singularity free black hole (BH) solutions have been proposed in conformally
invariant theories of gravity.

Latter on, conformal (Weyl) gravity characterized by a pure Weyl squared action
has taken a large amount of curiosity as an alternate theory to Einstein gravity.
Every conformal class of the Einstein solution occurs genuinely as a solution to the conformal gravity
by the equation of motion and the correspondence of conformal gravity. Mainly, in case of
the Neumann boundary condition, conformal gravity can single out an Einstein solution \cite{1pv, 1pw}.
Further, it has been investigated that unlike Einstein gravity, conformal gravity is
perturbatively renormalizable in four dimensions \cite{1px}, on basis of this aspect conformal gravity is more attractive
alternative to quantum gravity \cite{1py}.

Another interesting aspect of conformal gravity appears from cosmology.
It is an admitted fact that Einstein5 gravity can explain the physics within
the scale of the Solar system satisfactorily, but there are still several unsolved puzzles when tested to a
larger scale, such as the disparity with the observations of accelerating universe and
galactic rotation curves. Accordingly, anonymous entities, namely dark matter
and dark energy, have been proposed to remove the inconsistency issues.
However, one may wonder certainly wether it is feasible to modify the nature of gravity to illustrate
the physics at a larger scale, while maintaining the appropriate behaviors at the scale of
the Solar system. Subsequently conformal gravity permit more general solutions than Einstein gravity.
It can yield the effective potential compatible with the observed phenomenon,
which designate it as a fascinating modified theory of gravity \cite{1py1}-\cite{1py4}.

Furthermore, conformal theory and Einstein gravity may share similar metric solutions, while
thermodynamic quantities of the BH, such as the mass and entropy depends on the action of the theory
rather than the metric. The thermodynamical feature of the
conformal theory BH has been investigated in the four-dimensional spacetime in Refs.\cite{1pz1}-\cite{1pz4}.
In four-dimensions, there is an important feature that any conformal Einstein metric is
automatically a solution of conformal gravity.
In this respect, the light rays and test particle orbits can promote the understanding of
the physical properties of conformal field equations. Likewise, the motion of light
and fluid can be used to classify the conformal spacetime and to highlight its features.
The study of geodesic motion provides some information about the spacetime characteristics
of the BHs \cite{2}. The solutions of the geodesic equation of motion with the same mathematical
structure but different spacetime has been discussed in \cite{2a,2b}. The Jacobi inversion issues
have been solved \cite{2c,2d} with Schwarzschild anti de-Sitter spacetime and next studied in \cite{2e,2f}.
%The study of spacetimes $BTZ$, GMGHS BHs, cylindrical conformal and rotating BH and $f(R)$ gravity,
%is the great motivation in conformal gravity theory in a series of the papers \cite{2g}-\cite{2m}.
Some authors investigated the physical meaningful study of conformal gravity theory
\cite{3}-\cite{5}. The work for on the geodesics study in this theory have been discussed in \cite{6}.
Sultana et al. \cite{7} have studied the applications of geodesic solutions in the theory of conformal gravity.

The effects of electromagnetic fields around compact stars in conformal gravity have been explored by Toshmatov et al. \cite{7u}.
The observational test of conformal gravity has been done by using X-ray observations of supermassive BHs Ref.\cite{7v}.
Also, Toshmatov et al. \cite{7w, 7x} have discussed explicitly, the energy conditions and scalar perturbations around BHs in conformal gravity.
Recently, Haydarov et al. \cite{7y} have studied the motion of magnetized particle around BHs
in conformal gravity near the external magnetic fields.
In view of the above advancements in conformal theory of gravity, there is a blank space in the literature that no work has been done for the accretion onto BH
in this theory and we are capable to honestly fill this gap in the present paper.
For this, we plan to investigate spherically symmetric accretion onto BH
in conformal theory of gravity.

In astronomical observations, accretion causes to increase the mass of gravitating bodies
and also the angular momentum as well.
The accretion is used to explain the inflow matter towards
the central gravitating bodies. It has the significant status and a very special importance in
the astronomy and the cosmology. The most of the astrophysical bodies increase substantially
in mass due to accretion. The determination of mass accretion rate for astrophysical bodies
can provide a strong evidence for the presence of compact objects.
In general, the mass of gravitating bodies increases due to accretion onto it, but when phantom fluid accretes onto a gravitating object then its mass decreased \cite{8,9}.

The accretion onto celestial objects was started
by Hoyle and Lyttleton \cite{10}. Later on, the study of Hoyle and Lyttleton was
extended by Bondi and  Hoyle \cite{11} for gaseous dust cloud falling onto a compact object.
The transonic-hydrodynamics accretion for adiabatic fluid
onto gravitating object was investigated by Bondi \cite{12}
in Newtonian gravity, he found the existence of only one saddle-type sonic point in an adiabatic flow .
%After that a comprehensive literature has been dedicated to the
%observational and theoretical studies of accretion \cite{13,14} in Einstein theory of gravity.
The general relativistic
model of spherically symmetrically accretion onto Schwarzschild BH
was investigated by Michel \cite{15} that has been further extended by many authors \cite{13}-\cite{19}.
%In modern cosmology, the dark energy is the expressive debate
%with negative pressure \cite{20}.

It has been observed \cite{21}, that the theory of GR is facing a large number
of challenges that has led to the introduction of the dark matter
and dark energy with ordinary matter \cite{22}. Accretion of BHs
in the presence of dark energy studied has been studied by Babichev et al \cite{24}. They have
examined the behavior of different types of dark energy in the locality
of Schwarzschild and electrically charged BHs.
Debnath \cite{25} has studied the Babichev et al. \cite{24} and proposed a study of spherically
symmetrically accretion onto some generalized BHs. Moreover, accretion
of phantom fluid onto BHs has been formulated in \cite{26}-\cite{28}.

The BHs, planets and stars are
formed due to accretion during the evolutionary phases of dusty plasma clouds.
The existence of supermassive BHs in the center of galaxies intimate that
these BHs could have been developed through the accretion.
%The accretion disk does into existence because of acceleration of gases and dust
%around extensive objects. Accreting gas has various important issues,
%the presence of an event horizon is one of those
%which accreting gas vanishes and few radiations are emitted near BHs during the process.
The accretion onto higher dimensional BHs \cite{29} and string cloud model \cite{30} has been investigated in detail and it has been predicted that the matter accretion rate decreases with the increase of dimension
of spacetime. It has been shown \cite{30} that  accretion rate increases by the
increase of the string cloud parameter.
Furthermore, many authors \cite{31}-\cite{34} have analyzed the radial flows of perfect fluids
and dark energy onto some BHs in modified theories of gravity.
Sharif and his collaborators  \cite{35}-\cite{41} have studied the phantom accretion onto charge de-Sitter and string
BHs.

Another approach to the BH accretion comes from the modified theories of
gravities where, several authors have investigated the radial flows on static BHs \cite{41a},
quantum gravity corrections to accretion onto BH \cite{41b}, accretion onto a non-commutative BH \cite{41c}
and accretion onto BH in scalar-tensor-vector gravity \cite{41d}.
The accretion of cyclic and heteroclinic fluid flows
near $f(R)$ and $f(T)$ theories BHs have been explored in \cite{42,43}. The Hamiltonian approach,
has been used by several authors \cite{44}-\cite{49} to determine the maximum accretion flows of perfect fluid.

The purpose of the present research is to study the spherically symmetrically fluid
accretion onto conformal gravity BH.
We have analyzed the accretion process by applying the Hamiltonian approach (as developed in \cite{48,49}).
The accretion of various forms of fluids (such
as ultra-relativistic fluid, radiation fluid, ultra-stiff fluid and sub-relativistic fluid) onto conformal gravity BH
would provide the possibility of testing the conformal theory in strong gravity regimes.
From the effects of conformal gravity BH parameters $\gamma$, $k$ and
$\beta$, we have investigated that the consequences of fluid accretion in
case of conformal gravity BH are more significant and metric of the
conformal gravity BH is more generalized as compared to BH metrics available in literature in the context of accretion flows.

The setting of this research is as follows: The conformal gravity BH solution
with different values of parameters is presented in section \textbf{2}.
In section \textbf{3}, we have formulated the horizon structure of conformal gravity BH.
Section \textbf{4} is dedicated to calculate the conservation laws and general formulism for accretion onto conformal gravity BH.
The sound speed at the critical points is formulated in section \textbf{5}. Further, in the subsequent sections,
we have applied the Hamiltonian approach for the accretion with isothermal equation of state.
In section \textbf{6}, we have analyzed the nature of flow around conformal gravity BH for
various cases of fluids such as ultra-stiff fluid, ultra-relativistic fluid, radiation fluid and sub-relativistic fluid.
The polytropic fluid accretion has been discussed explicitly in section \textbf{7}.
The mass accretion rate for various BHs
is formulated in section \textbf{8}. Finally, we concluded the results of the paper in section \textbf{8}.
Here, we have used the signature of metric as $(-,+,+,+)$ and the geometric units $G=c=1$.

\section{Conformal Gravity Black Hole}
According to conformally invariant theory of gravity, the action is invariant
with respect to general conformal transformations and coordinate transformations.
Let a regular quantity be singular in a reference frame but not in alternative one due to a conformal
transformation, the singularity is not physical but exact an artefact of the reference frame.
Here, we have a \textit{conformal singularity}, which is related to the preference of the
conformal factor but not an intrinsic singularity of spacetime.
It is noted that one cannot apply the same mathematical mechanism for studying spacetime
singularities in conformal gravity and Einstein gravity. For instance, the Kretschmann scalar
and the scalar curvature are not invariant with respect to conformal transformations, so they are not
linked with any intrinsic property in conformal gravity spacetime. %The intension of geodesics is also
%dissimilar in conformal gravity due to massive particles which are not allowed because they are not
%agreeable with conformal gravity.
Therefore, the general action of conformal theory can be developed by on the basis of following
four points.
\begin{itemize}
  \item  It is a completely covariant advancement theory of General Relativity.
  \item  It is an additional symmetry principle or local conformal invariance and the existence
  of the symmetry principle prevents the Einstein Hilbert action and cosmological term in the action.
  \item  The conformal transformation of the theory is $g_{\mu\nu}\rightarrow \Omega^2(x)g_{\mu\nu}$.
  \item  The conformal gravity action is defined in terms of Weyl tensor $C_{\eta\lambda\mu\nu}$
  and a coupling constant $\alpha_g$, which is a dimensionless constant and this allows the conformal gravity
  theory is a quantum theory of gravity. The action of conformal gravity contributes to fourth
  order of equation of motion in the presence of ghosts. The fourth order equations of motion involve
  more constants of integration and also solutions contain more parameters.
\end{itemize}
The action and the field equations of metric are given by
\begin{eqnarray}
S_{CG}&=&-\alpha_g\int d^4 x(-g)^{1/2}C_{\eta\lambda\mu\nu}C^{\eta\lambda\mu\nu},\label{1}\\
C_{\eta\lambda\mu\nu}&=&R_{\eta\lambda\mu\nu}-\frac{1}{2}(g_{\eta\mu} R_{\lambda\nu}-g_{\eta\nu} R_{\lambda\mu}+g_{\lambda\nu}R_{\eta\mu}-g_{\lambda\mu}R_{\eta\nu})+\frac{R}{6}(g_{\eta\mu}g_{\lambda\nu}-g_{\eta\nu}g_{\lambda\mu}),\label{2}
\end{eqnarray}
where $\alpha_g$ is the pure dimensionless constant and $C_{\eta\lambda\mu\nu}$ represents the Weyl tensor.
The following gravitational field equations are achieved by varying the action (1) in the
presence of $W_{\mu\nu}$ and the energy momentum tensor $T_{\mu\nu}$ given by
\begin{eqnarray}
2\alpha_g W_{\mu\nu}=&&\frac{1}{2}T_{\mu\nu}.\label{3}\\
W_{\mu\nu}=&&\frac{1}{3}\nabla_{\mu}\nabla_{\nu}R-\nabla_{\lambda}\nabla^{\lambda}R_{\mu\nu}+\frac{1}{6}(R^2+\nabla_{\lambda}\nabla^{\lambda}R
\\\nonumber&&-3R_{\eta\lambda}R^{\eta\lambda})g_{\mu\nu}+2R^{\eta\lambda}R_{\mu\eta\nu\lambda}-\frac{2}{3}RR_{\mu\nu}.\label{4}
\end{eqnarray}

Therefore, the exact vacuum solution is given by the metric
\begin{eqnarray}
ds^{2}=&&-\left(1-\frac{\beta(2-3\beta\gamma)}{r}-3\beta\gamma+\gamma r-kr^2\right)dt^{2}+
\\\nonumber&&\left(1-\frac{\beta(2-3\beta\gamma)}{r}-3\beta\gamma+\gamma r-kr^2\right)^{-1}dr^{2}+r^{2}(d\theta^{2}+ \sin^{2}\theta d\phi^{2}).\label{5}\\
\end{eqnarray}
The above metric can be written as
\begin{eqnarray}
ds^{2}&&=-A(r)dt^{2}+\frac{1}{A(r)}dr^{2}+r^{2}(d\theta^{2}+ \sin^{2}\theta d\phi^{2}),\label{7}\\
A(r)=&&1-\frac{\beta(2-3\beta\gamma)}{r}-3\beta\gamma+\gamma r-kr^2.\label{6}
\end{eqnarray}

Here $A(r)>0$, while $\beta$, $\gamma$ and $k$ are constants of integration.
The choice $\gamma=0$ yields the Schwarzschild-de Sitter solution
and $\gamma=k=0$ the Schwarzschild solution.

If all three parameters of BH $\gamma$, $\beta$ and $k$ are not equal to zero,
the BH can be termed as case \textbf{1}, which is a general case.
If we impose $\gamma=0$, $A(r)=1-\frac{2\beta}{r}-kr^2$ (Schwarzschild-de Sitter solution),
which is the case \textbf{2}. We obtained the Schwarzschild solution $A(r)=1-\frac{2\beta}{r}$ for $\gamma=k=0$,
which is the case \textbf{3}. In all cases $\beta$ is considered
as mass of BH and $k$ behaves as cosmological constant \cite{1py1, 1py2}.

\section{Horizon structure}
In this section, we study the horizon structure of the conformal gravity theory BH.
For the metric function $A(r)$ given in Eq. (\ref{6}), the effect of the
parameters $\beta$, $\gamma$ and $k$ is significent. It is well known that
the conformal gravity BH appears to describe a massive body fixed in a
conformally flat space. The conformally flatness of the spherically symmetric space
is characterized by the absolute influence of the massive body. As aforementioned by the analysis
of Weyl tensor, the shattering of conformally flatness is evident at infinity.
In view of the above arguments, the conformal theory BH has the following structure:
\begin{itemize}
  \item When one take the mass function $\beta=0$, the conformally flat solution is
  obtained \cite{1py1}.
  \item When $\beta\neq0$, the coefficient of the $\frac{1}{r}$ term disappear as $r$ approaches to infinity.
  The non-vanishing contribution exists due to the constant term $-3\beta\gamma$.
  \item For the line element, the influence of the Newtonian term $\frac{1}{r}$,
  which should reign at small lengths with $\gamma r$ term entity the further dominant
  one at larger lengths \cite {1py1}.
\end{itemize}

\begin{figure}
\begin{center}
\includegraphics[width=80mm]{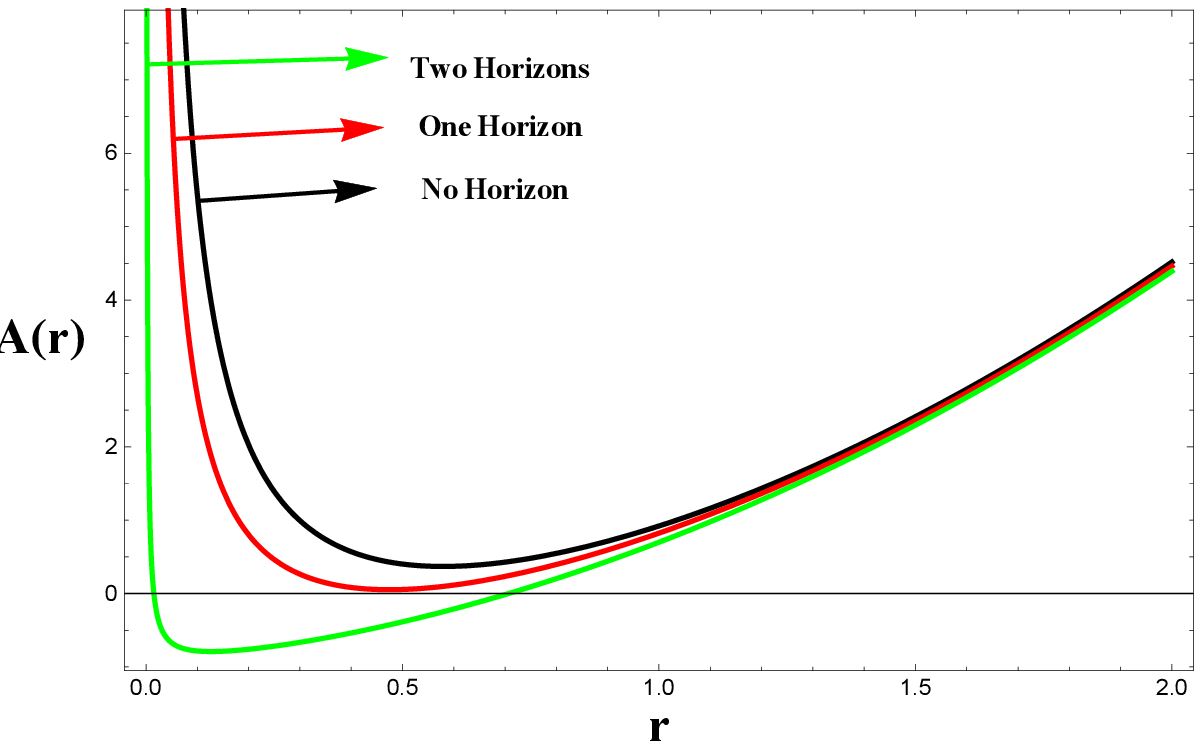}
\includegraphics[width=80mm]{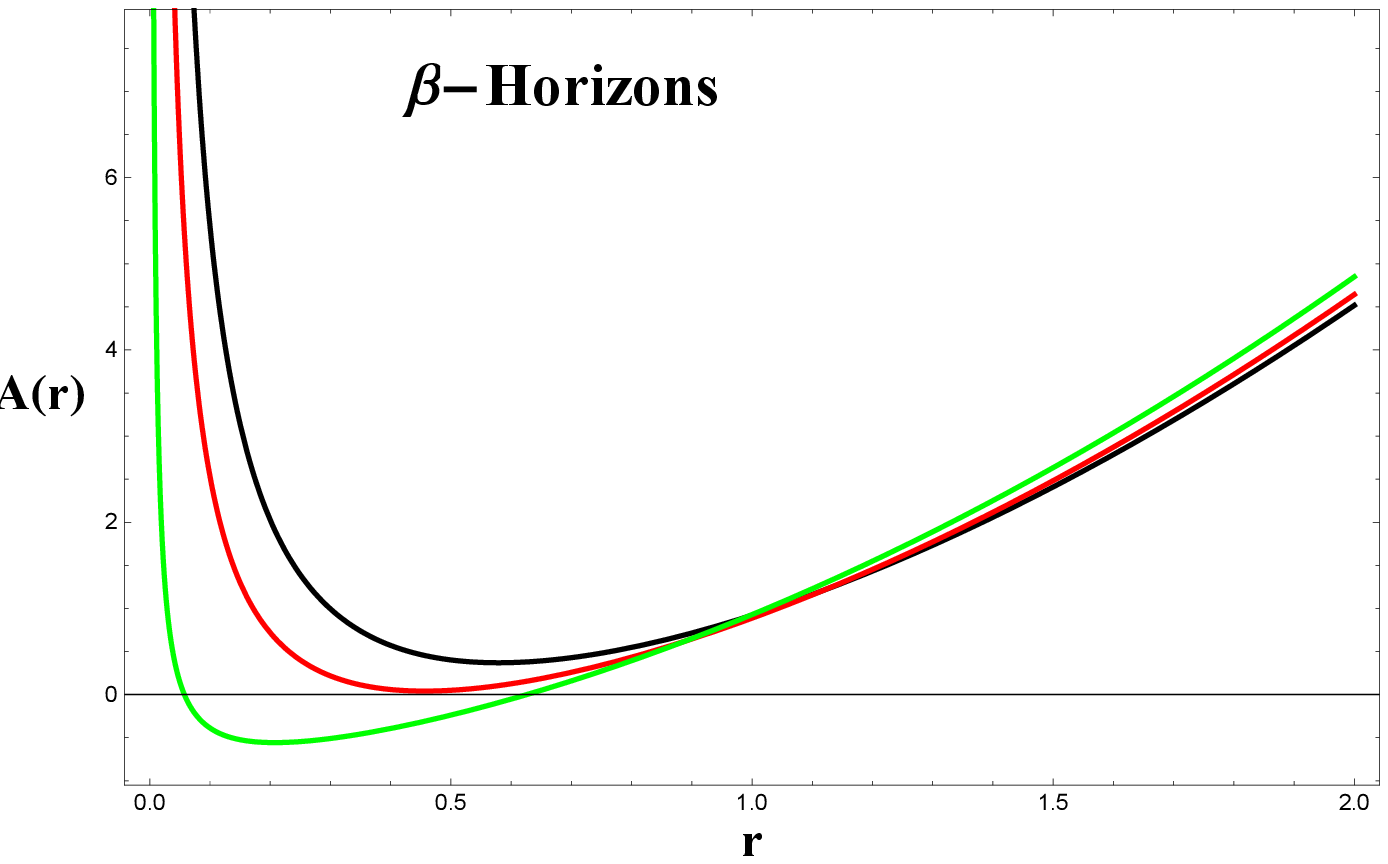}
\includegraphics[width=80mm]{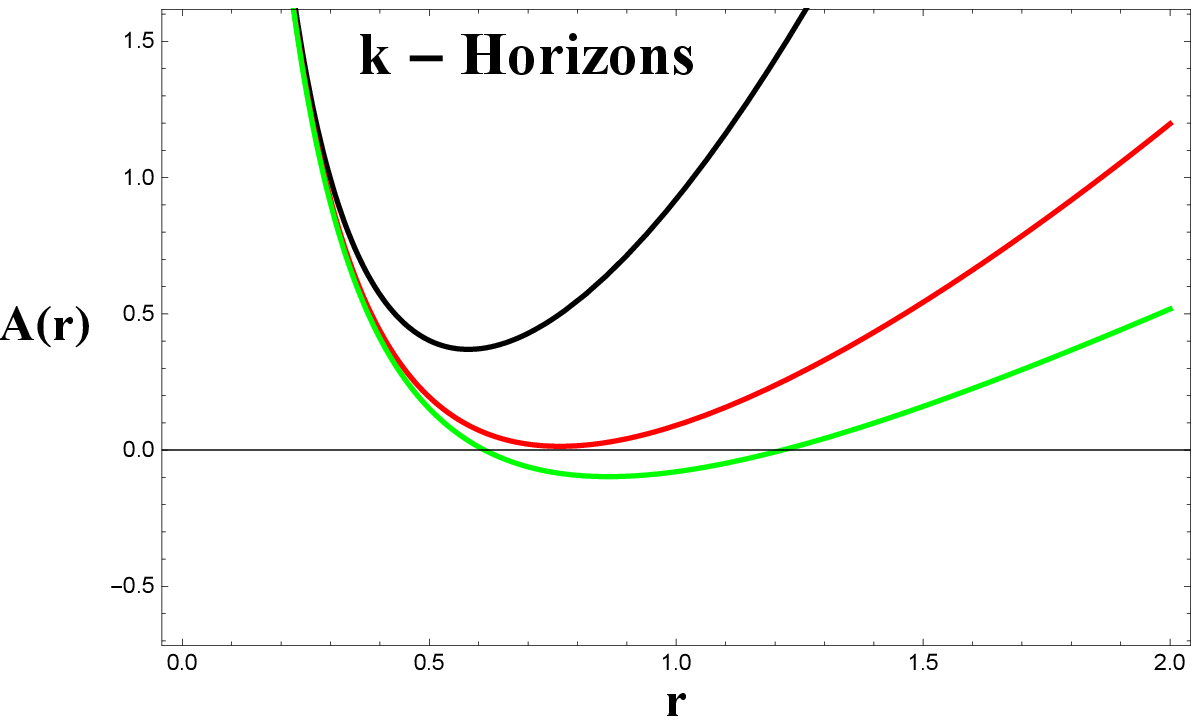}
\caption{The horizon structure of conformal BH displays the behavior of $A(r)$ versus $r$. Left panel for
various values of $\gamma$ and others parameters are taken as fixed $k=-1$, $\beta=0.96$. In the right panel,
$\beta$ horizons mean, we take various values of $\beta$. In the bottom panel,
$k$ horizons mean, we take various values of $k$. (Figure color online).}
\end{center}
\end{figure}

\begin{figure}
\begin{center}
\includegraphics[width=80mm]{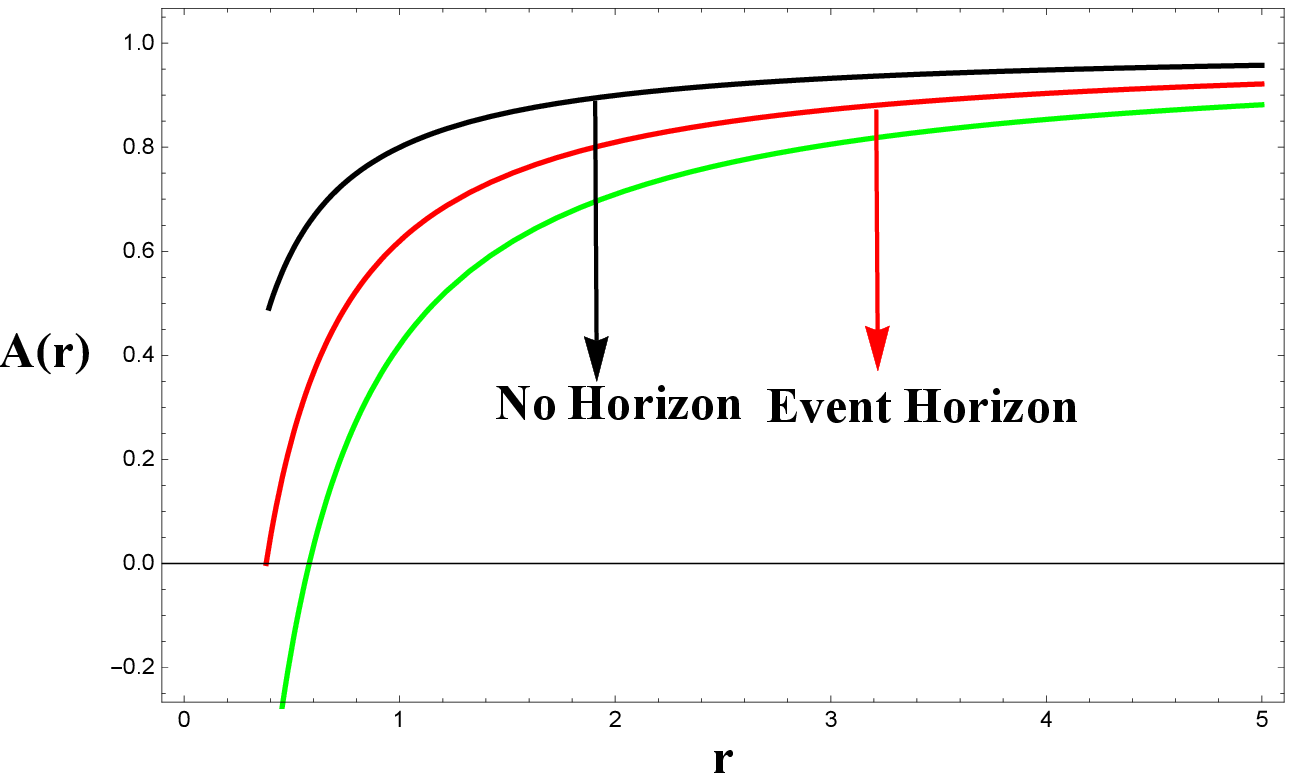}
\includegraphics[width=80mm]{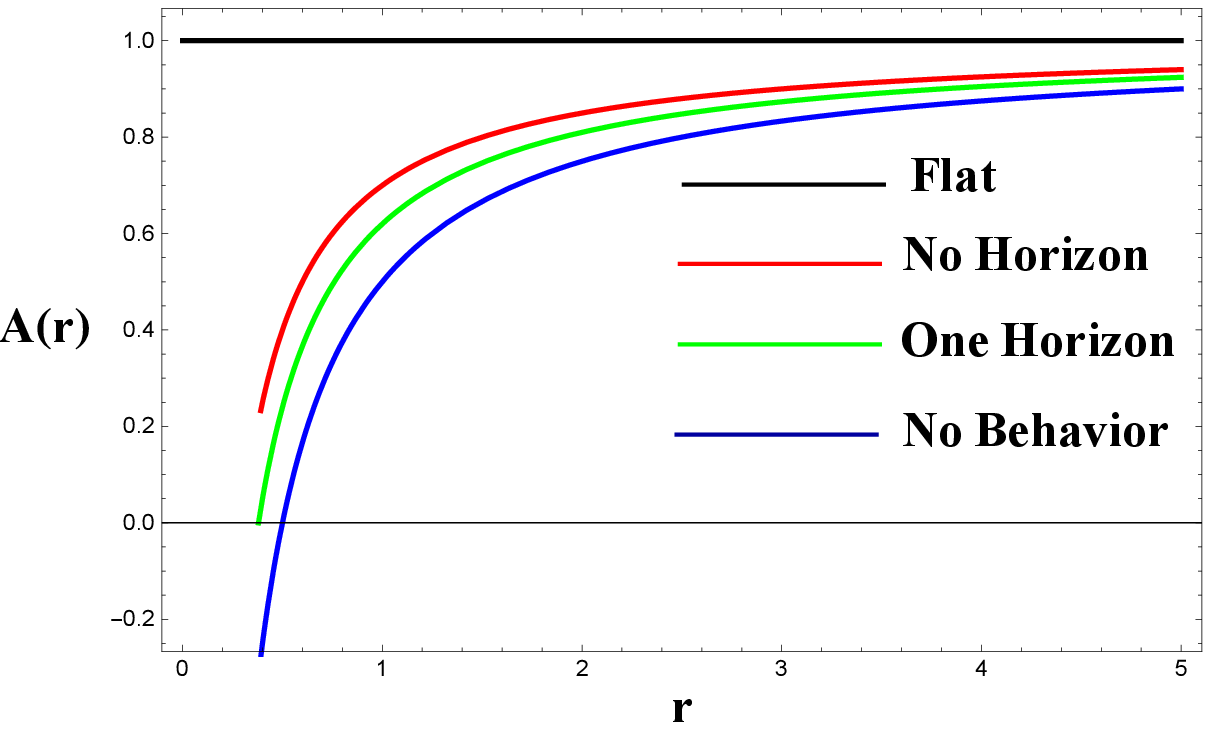}
\caption{The horizon structure of Schwarzschild-de Sitter BH (left panel) and Schwarzschild BH (right panel)
display the behavior of $A(r)$ versus $r$.
The horizon curves of Schwarzschild-de Sitter BH are obtained for $\beta=0.10, 0.19, 0.29$ and other parameters are taken as
fixed $k=-0.0001$, $\gamma=0$. The horizon curves of Schwarzschild BH are obtained for $\beta=0, 0.15, 0.19, 0.25$ and others parameters are taken as
fixed $k=0$, $\gamma=0$. (Figure color online).}
\end{center}
\end{figure}

In the radial distance $g_{rr}=0$ or $A(r)=0$, one would have an event horizon.
%The rule of an event horizon in the geometry is a
%radial distance from the center of the core where the metric is singular, without the intrinsic
%singularity which cannot be isolated through coordinated transformation.
Figure \textbf{1} (left panel) has the following key points:
\begin{itemize}
  \item We have two horizons (green curve) for $\gamma=0.70$.
  \item We have one horizon (red curve) for $\gamma=0.84$.
  \item We have no horizons (black curve) for $\gamma=0.95$.
\end{itemize}
Figure \textbf{1} (right panel) has the following key points:
\begin{itemize}
  \item We have two horizons (green curve) for $\beta=0.73$.
  \item We have one horizon (red curve) for $\beta=0.86$.
  \item We have no horizons (black curve) for $\beta=0.96$.
\end{itemize}
Figure \textbf{1} (bottom panel) has the following key points:
\begin{itemize}
  \item We have two horizons (green curve) for $k=-0.0001$.
  \item We have one horizon (red curve) for $k=-0.17$.
  \item We have no horizons (black curve) for $k=-1$.
\end{itemize}

Figure \textbf{2} left panel presents the Schwarzschild-de Sitter BH horizon structure for $\gamma=0$,
there exists only one event horizon with the variations of $\beta$.
According to the right panel of Fig. \textbf{2}, one gets Schwarzschild solution for $\gamma=0$ and $k=0$,
there is only one event horizon.
When we take $\beta=0$, the Schwarzschild solution is flat (see black curve).
It is noted that when all parameters appearing from conformal gravity are set equal to zero
then the solution is conformally flat \cite{1py1}.

\section{Spherically Symmetrically Accretion}
The study of spherical accretion onto BHs
describes the movement of fluid near the event horizon.
We assume two important laws that fully characterized
the spherically accretion of perfect isotropic fluid.
One is the law of conservation of mass and
the other is the law of conservation of energy.
We begin by the equation of continuity defined by Rezzolla and
Zanotti \cite{55} in the following form
\begin{equation}
\nabla_\mu J^\mu =0,\label{8}
\end{equation}
where $J^\mu=nu^\mu$ and $n$ is the proper baryon number density and
$u^\mu=\frac{dx^\mu}{d\tau}$ is the velocity-fluid.
Now, we introduce the matter the energy-momentum tensor as $T^{\mu\nu} =(\varrho+p)u^\mu u^\nu+pg^{\mu \nu}$,
with $\varrho$ as the energy density, $p$ is the pressure and
$u^\mu=\frac{dx^\mu}{d\tau}=(u^0,u^1,0,0)$ is $4$-velocity of the particle.
The conservation of energy momentum for perfect fluid is given by
\begin{equation}
\nabla_\mu T^{\mu \nu} =0.\label{9}
\end{equation}
The fluid follows the normalization condition $ u^\mu u_\mu = -1$, which leads to
\begin{equation}
u^0 =\frac{\sqrt{A(r)+u^2}}{A(r)},\label{10}
\end{equation}
also
\begin{equation}
u_0 =- \sqrt{A(r)+u^2}.\label{11}
\end{equation}
Using the equatorial plane $(\theta=\pi/2)$, the mass conservation Eq.(\ref{8}) can be written as
\begin{equation}
\frac{1}{r^2}\frac{d}{dr}(r^2 nu)=0,\label{12}
\end{equation}
after integrating, it gives
\begin{equation}
r^2 nu=c_1,\label{13}
\end{equation}
where, $c_1$ is an integration constant. For accretion, the fluid velocity is $u^r<0$
and therefore $c_1<0$ in above equation.
We define the enthalpy as $h(\varrho,p,n)=\frac{\varrho+p}{n}$.
For smooth flow, Eq. (\ref{9}) can be written as
\begin{equation}
nu^\mu \nabla_\mu (hu^\nu)+g^{\mu\nu} \partial_\mu p=0.\label{14}
\end{equation}
Moreover, if the entropy of a moving fluid along a streamline is constant,
then the fluid flow should be an isentropic \cite{42}.
So, above equation reduces to
\begin{equation}
u^\mu \nabla_\mu (hu^\nu)+\partial_\nu h=0.\label{15}
\end{equation}
The zeroth component of above equation yields
\begin{equation}
\partial_r (hu_0)=0,\label{16}
\end{equation}
after integrating above equation gives
\begin{equation}
h\sqrt{A(r)+(u)^2}=c_2,\label{17}
\end{equation}
where $c_2$ is constant of integration. Thus, Eqs. (\ref{13}) and (\ref{17})
are important for the critical flow of fluid onto the considered BH.
\section{Hamiltonian dynamical approach}
\subsection{Sound Speed at Sonic Points}
The sonic point (critical point) is a point, where the velocity of the moving gas
must be equal to the local sound speed. In view of this definition, the maximum accretion rate occurs,
when fluid passes through the critical point. Here, we are interested to calculate the critical points of the flow
and the sound speed at these points, for this, we assume the barotropic fluid
with constant enthalpy that is $h=h(n)$.
Therefore, the equation of state for this flow becomes \cite{56}
\begin{equation}
\frac{dh}{h} = a^2\frac{dn}{n},\label{18}
\end{equation}
where $a$ represents the local sound speed. Also, above equation gives $\ln h=a^2\ln n$.
Using Eqs.(\ref{10}), (\ref{17}) and (\ref{18}), we get
\begin{equation}
\left[\left(\frac{u}{u_0}\right)^2-a^2\right](\ln u)_{,r}=\frac{1}{r(u_0)^2}\left[2a^2(u_0)^2-\frac{1}{2}r A'(r)\right].\label{19}
\end{equation}
Now, for critical points, both sides of the above equation
must be equal to zero. So, the sound speed at the critical point becomes
\begin{equation}
a^2_c = \left(\frac{u_c}{u_{0c}}\right)^2,\label{20}
\end{equation}
where the quantities, $a_c$ , $r_c$ and $u_c$ designate the sound speed, distance
and velocity of the fluid at the critical point, respectively.
Therefore, another result of Eq. (\ref{19}) at the critical point is given by
\begin{equation}
2a^2_c(u_{0c})^2-\frac{1}{2}r_cA'_{rc} = 0.\label{21}
\end{equation}
From Eqs.(\ref{20}) and (\ref{21}), one can get the radial velocity at the critical points, which is given by
\begin{equation}
(u_c)^2 = \frac{1}{4}r_cA'_{rc}.\label{22}
\end{equation}
Using Eqs. (\ref{11}), (\ref{21}) and (\ref{22}), we obtain
\begin{equation}
r_cA'_{rc} = 4a^2_c[A(r_c)+(u_c)^2].\label{23}
\end{equation}
Thus, the sound speed is given by
\begin{equation}
a^2_c = \frac{r_c A'_{rc}}{r_cA'_{r_c}+4A(r_c)}.\label{24}
\end{equation}
Thus, one can obtain the critical points as $(r_c,\pm u_c)$ using
Eqs. (\ref{22}) and (\ref{24}), if one has the value of sound speed.
\subsection{Isothermal Test Fluids}
At very high speed, the fluid does not transfer its the heat to the surrounding, so it is the the adiabatic situation
, we introduce $a^2=\frac{dp}{d\varrho}$
to solve analytically the equations of motion of the fluid.
For the analytical solutions, we define an important equation of state $p=\omega\varrho$ with
the energy density $\varrho$ and equation of the state parameter $\omega$. Here, $0<\omega\leq1$ as constrained in \cite{32}.
After combining $p=\omega\varrho$ with $a^2=\frac{dp}{d\varrho}$,
one gets, $a^2=\omega$. Applying the first law of thermodynamics
\begin{equation}
 \frac{d\varrho}{dn}=\frac{\varrho+p}{n}=h.\label{25}
\end{equation}
Integrating the above equation from the critical point to any point inside the fluid, we get
\begin{equation}
n=n_c \exp\left(\int^{\varrho}_{\varrho_c}\frac{d\varrho'}{\varrho'+p(\varrho')}\right).\label{26}
\end{equation}
With the help of $p=\omega\varrho$, the above Eq. (\ref{26}) gives
\begin{equation}
n=n_c \left(\frac{\varrho}{\varrho_c}\right)^\frac{1}{\omega+1}.\label{27}
\end{equation}
Using enthalpy $h(\varrho,p,n)=\frac{\varrho+p}{n}$ in the above relation, we obtain
\begin{eqnarray}
h=\frac{(\omega+1)\varrho_c}{n_c}\left(\frac{n}{n_c}\right)^\omega.\label{28}
\end{eqnarray}
The following relation with constant of integration
By using Eqs. (\ref{28}) and (\ref{17}), we get
\begin{equation}
n^\omega\sqrt{A(r)+(u)^2}=c_3.\label{29}
\end{equation}
where $c_3=\frac{C_2n^{1-\omega}_c}{(\omega+1)\varrho_c}.$
Combining Eqs.(\ref{25}) and (\ref{10}), we get
\begin{equation}
\sqrt{A(r)+(u)^2}=C_3r^{2\omega}(u)^\omega.\label{30}
\end{equation}
The Hamiltonian can be defined as \cite{42,43}
\begin{equation}
H=\frac{A^{1-\omega}}{(1-v^2)^{1-\omega}v^{2\omega}r^{4\omega}}.\label{31}
\end{equation}
where $v \equiv \frac{dr}{fdt}$, is the three-dimensional speed for the
radial motion of particle in the equatorial plane.
Thus, we have
\begin{equation}
v^2=\left(\frac{u}{fu^0}\right)^2=\frac{u^2}{u^2_0}=\frac{u^2}{f+u^2}.\label{32}
\end{equation}
Further, by using (\ref{22}) and (\ref{23}) the critical points can be obtained as follows
\begin{eqnarray}
(u_c)^2&=&\frac{1}{4}r_c A'_c,\label{33}\\
(u_c)^2 &=&\omega\left(\frac{1}{4}r_c A'_c+A_c\right).\label{34}
\end{eqnarray}
Consequently, for the classification of fluid flow, the generalized expression
(\ref{31}) can be solved numerically by choosing the any value of $\omega$ satisfying $0<\omega\leq1$.
In the coming subsections, we have assumed the four kinds of fluid such as, ultra-stiff fluid, ultra-relativistic fluid,
radiation fluid and sub-relativistic fluid.
\section{Analysis of Various Cases of Black Holes}
\subsection{Case \textbf{1}.}
The first and general case of the line element is a conformal BH, which is given by Eq.(\ref{6}).
%\begin{equation}
%A(r)=1-\frac{\beta(2-3\beta\gamma)}{r}-3\beta\gamma+\gamma r-kr^2.\label{35}
%\end{equation}
There are four subcases of case \textbf{1}:
\begin{enumerate}
\item [\textbf{6.1.1}]\underline{\textbf{Hamiltonian for ultra-stiff fluid  ($\omega=1$):}}\\
In this case, we have $p=\varrho$, from the equation of state
and critical point and event horizon are equal,
that is $r_h=r_c$, with the condition $A_c=0$, which can be easily obtained by
using Eqs. (\ref{33}) and (\ref{34}).
The Hamiltonian (\ref{31}) for this type of fluid takes the form
\begin{equation}
H=\frac{1}{v^2{r_c}^4}.\label{36}
\end{equation}
\item [\textbf{6.1.2}]\underline{\textbf{
{Hamiltonian for ultra-relativistic fluid  ($\omega=1/2$):}}}\\
In ultra-relativistic fluids, the relation between energy density ($\varrho$) and pressure ($p$)
is $p=\varrho/2$, with $\omega=1/2$ in the equation of state,
it means that the energy density is grater than the pressure in this case.
From Eqs. (\ref{33}) and (\ref{34}), we get $r_c A'_c-4A_c=0$, which gives
\begin{equation}
2k{r_c}^3-3\gamma {r_c}^2+4(3\beta\gamma-1){r_c}+5\beta(2-3\beta\gamma)=0.\label{37}
\end{equation}
The real solution of above equation is
\begin{eqnarray}
{{r_c}=\frac{\gamma }{2 k}-\frac{24 k-72 k \beta  \gamma +9 \gamma ^2}{{3\times 2^{2/3} k Q}}+
{\frac{Q}{6\times
2^{1/3} k}}},
\end{eqnarray}
 where
\begin{eqnarray}\nonumber
&&Q=\Big(-1080 k^2 \beta +216 k \gamma +1620 k^2 \beta
^2 \gamma -648 k \beta  \gamma ^2+54 \gamma ^3\nonumber\\&& +\sqrt{4 \left(-24 k+72 k \beta \gamma -9 \gamma ^2\right)^3+\left(-1080 k^2 \beta +216 k \gamma +1620
k^2 \beta ^2 \gamma -648 k \beta  \gamma ^2+54 \gamma ^3\right)^2}\Big)^{1/3}.\nonumber
\end{eqnarray}
Using $r_c$ from the above expression, we get $v_c$ from Eq.(\ref{33}) and
have two critical points as $(r_c,\pm {v_c})$.
The Hamiltonian (\ref{31}) reduces into the form:
\begin{equation}
H=\frac{\sqrt{A}}{{r_c}^2v \sqrt{1-v^2}}.\label{38}
\end{equation}
The graphical behavior can be seen between $v$ and $r_c$ with particular choice of $H=H_c$.
%From the above relation (\ref{38}), we get
%\begin{equation}
%v^2=\frac{1\pm \sqrt{1-4A(r)}}{2}.\label{39}
%\end{equation}
\item [\textbf{6.1.3}]\underline{\textbf{{Hamiltonian for radiation fluid  ($\omega=1/3$):}}}\\
For radiation fluids, the equation of state takes the form $p=\varrho/3)$.
Using Eqs. (\ref{33}) and (\ref{34}), we have $r_c A'_c-2A_c=0$, which leads to
\begin{equation}
\gamma {r_c}^2 -(3\beta\gamma-1){r_c}-3\beta(2-3\beta\gamma)=0.\label{40}
\end{equation}
The critical solutions are
\begin{equation}
r_{c_\pm}=\frac{2(3\beta\gamma-1)\pm2}{2\gamma}.\label{41}
\end{equation}
The Hamiltonian (\ref{31}) in this case is given by
\begin{equation}
H=\frac{A^{\frac{2}{3}}}{{r_c}^\frac{4}{3}v^{\frac{2}{3}}(1-v^{2})^{\frac{2}{3}}}.\label{42}
\end{equation}
\item [\textbf{6.1.4}]\underline{\textbf{{Hamiltonian for sub-relativistic fluid  ($\omega=1/4$):}}}\\
For such fluids, the equation of state obeys the form $p=\varrho/4$.
This form shows the energy density exceeds than the isotropic pressure.
Using Eqs. (\ref{33}) and (\ref{34}), we get $4A_c-3r_cA'_c=0$,
which reduces to
\begin{equation}
2k{r_c}^3+\gamma {r_c}^2-4(3\beta\gamma-1){r_c}-7\beta(2-3\beta\gamma)=0.\label{43}
\end{equation}

\begin{eqnarray}
{r_c}=\frac{\gamma }{3 k}-\frac{2^{1/3} \left(-3 k+9 k \beta  \gamma -\gamma ^2\right)}{3 k S}+
\frac{S}{3\ 2^{1/3} k}
\end{eqnarray}
$
S=\Big(P+\sqrt{4 \left(-3 k+9 k \beta  \gamma -\gamma ^2\right)^3+P^2}\Big)^{1/3}$\\
$P=-54 k^2 \beta +9 k \gamma +81 k^2
\beta ^2 \gamma -27 k \beta  \gamma ^2+2 \gamma ^3$\\
After determining the critical points, the Hamiltonian is given by
\begin{equation}
H=\frac{A^{\frac{3}{4}}}{{r_c}v^{\frac{1}{2}}(1-v^{2})^{\frac{3}{4}}}.\label{44}
\end{equation}
\end{enumerate}

\subsection{Case \textbf{2}}
The second case of the line element is known as Schwarzschild-de Sitter BH,
the Hamiltonian and critical points for such BH can be obtained by taking $\gamma=0$ in \textbf{Case 1}.

\subsection{Case \textbf{3}}
The third and final case of the line element is called Schwarzschild solution, the Hamiltonian and critical points for such BH can be obtained by taking $\gamma=k=0$ in \textbf{Case 1}.
\begin{figure}
\begin{center}
\includegraphics[width=80mm]{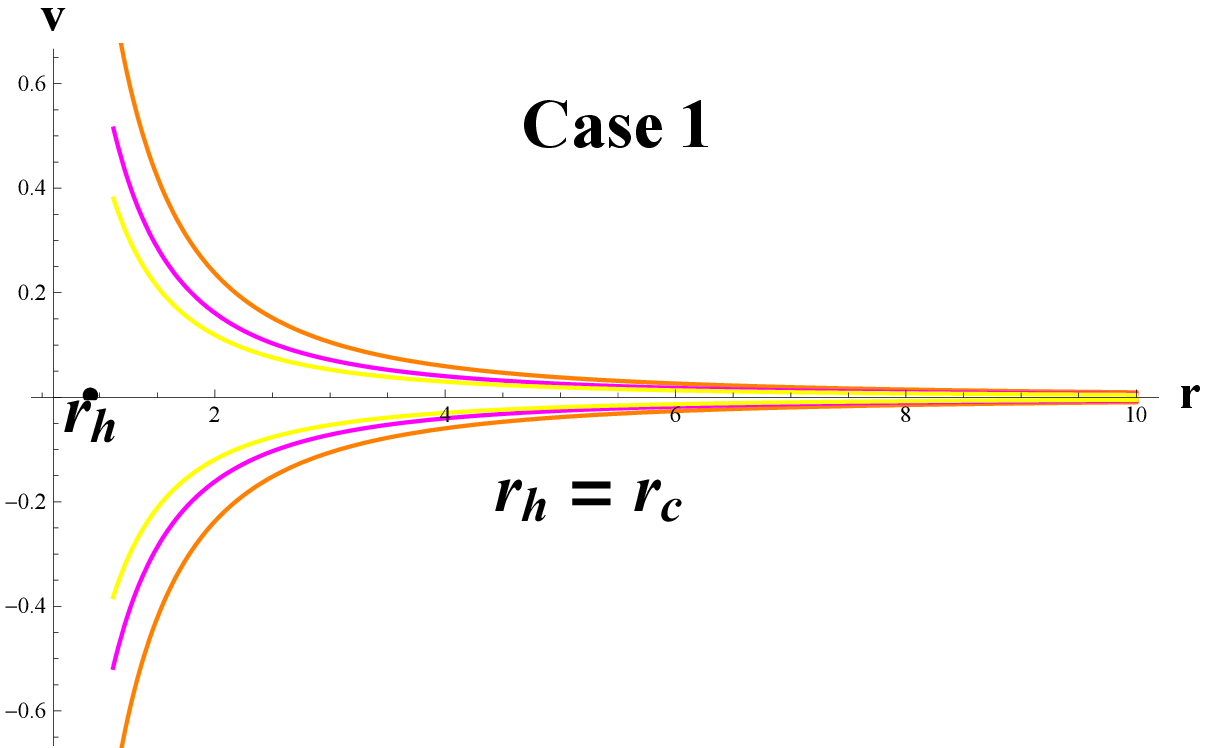}
\includegraphics[width=80mm]{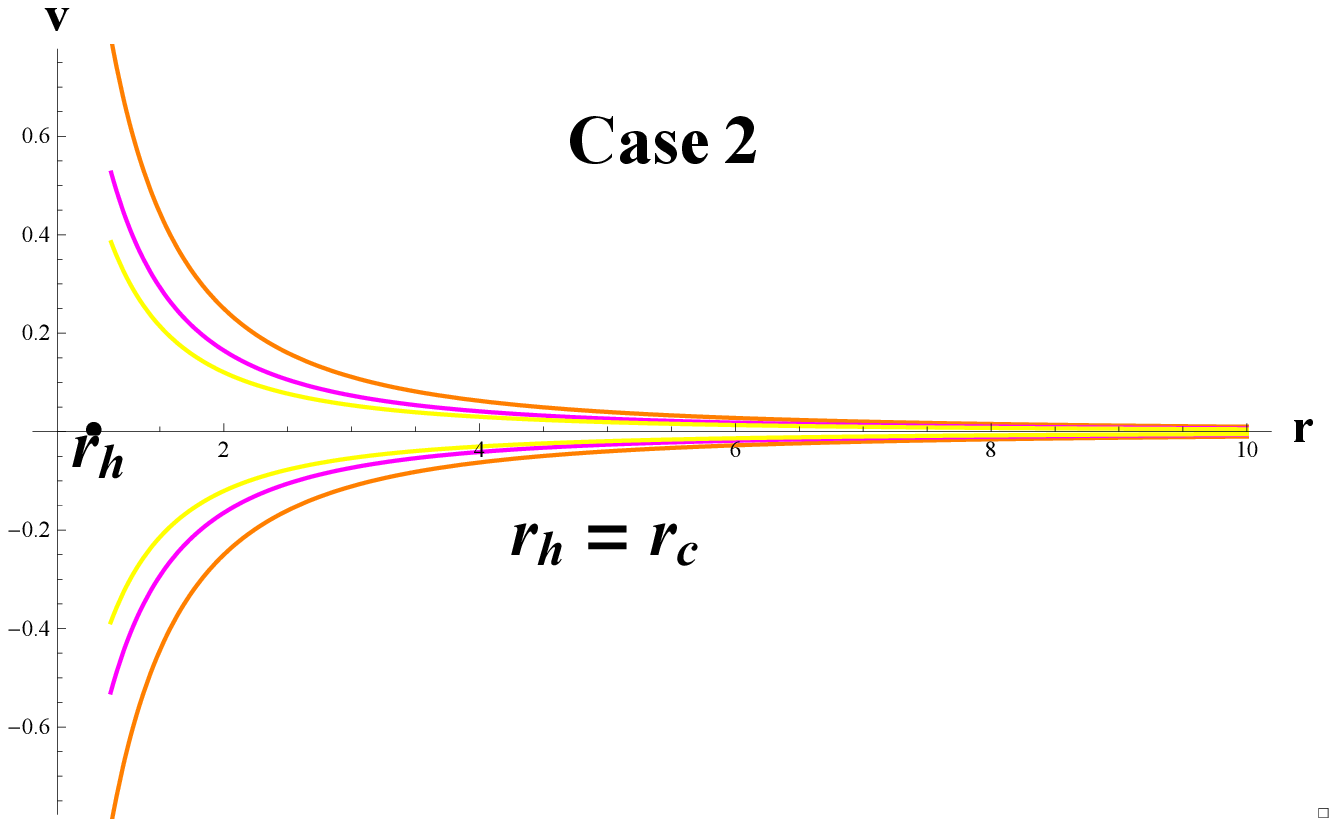}
\includegraphics[width=80mm]{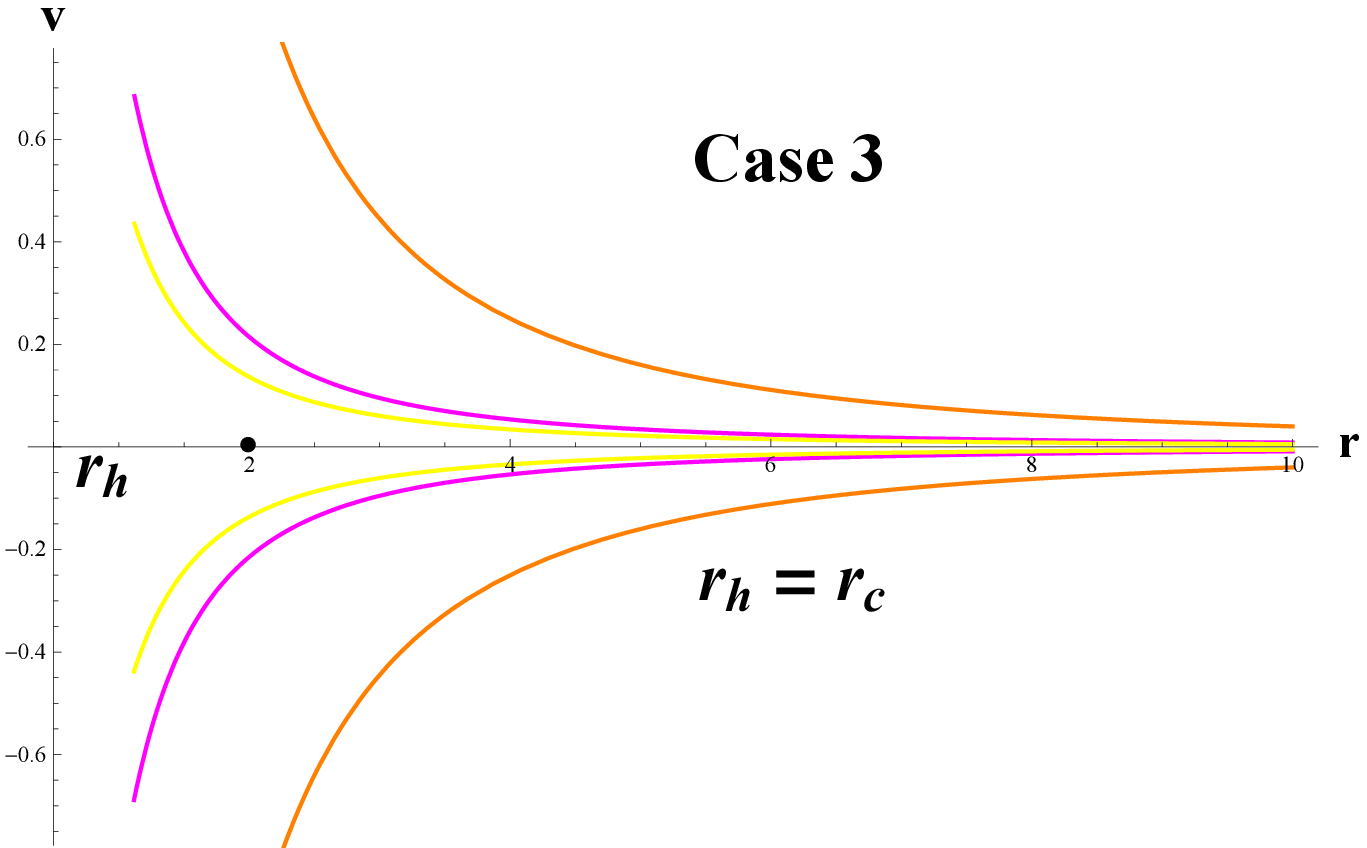}
\caption{The left panel (conformal gravity BH) displays the behavior of equation ($36$) with
conformal parameters $\gamma=0.1$, $k=-1$, $\beta=1$. The critical parameters are chosen as
$r_c\approx0.9738$, $v_c=1$, $H_c\approx1.112$. The right panel (Schwarzschild de-Sitter BH) displays the behavior of Eq.($36$)
with Schwarzschild de-Sitter BH parameters $\gamma=0$, $k=-1$, $\beta=1$. The critical parameters are chosen as
$r_c\approx1.001$, $v_c=1$, $H_c\approx1.001$. The bottom panel (Schwarzschild BH) displays the behavior of Eq.($36$) with Schwarzschild BH parameters
$\gamma=0$, $k=0$, $\beta=1$.  The critical parameters are chosen as $r_c=2$, $v_c=1$, $H_c\approx0.0625$. The representation of
colors for $H=H_c\rightarrow$ orange, $H>H_c\rightarrow$ magenta and yellow (Figure color online).}
\end{center}
\end{figure}
\subsection{Visualization of results for all cases}
This section elaborates the results of ultra-stiff fluids
($\omega=1$), ultra-relativistic fluids ($\omega=\frac{1}{2}$), radiation
fluids ($\omega=\frac{1}{3}$) and sub-relativistic fluids
($\omega=\frac{1}{4}$) for conformal gravity BH (Case 1),
Schwarzschild de-Sitter BH (Case 2) and Schwarzschild BH
(Case 3).
\begin{enumerate}
  \item \textbf{Ultra-stiff fluids ($\omega=1$):}
\newline
Figure \textbf{3} signifies the velocity $v$ of moving fluid versus the radius $r$
corresponding to various cases of BH.  The fluid motion occurs in the two regions, upper curves show
the region $v>0$ whereas the lower curves for the region $v<0$ for the three cases of BH.
It is noted here that the critical point is always equal to the critical horizon
for ultra-stiff fluid. Moreover, the critical points are closer in conformal gravity BH and Schwarzschild de-Sitter BH
as compare to the Schwarzschild BH. It is observed that the critical points are close to
the singularity in conformal gravity BH and Schwarzschild de-Sitter BH than the Schwarzschild BH.
  \item \textbf{Ultra-relativistic fluids ($\omega=\frac{1}{2}$):}
\newline
We see the velocity essence of moving fluid $v$ versus the radius $r$ for
aforementioned all three cases of conformal gravity BH by putting the
corresponding values of $A(r)$ in Fig. \textbf{5}. The critical
values of horizon, radius and velocity $(r_h,r_c,v_c)$ are nearly equal to
(0.974,~1.297,~0.707107), (1.3333,~1.3282,~0.707107) and
(2.0025,~2.524,~0.707107) for the cases $1$, $2$ and $3$, respectively. For
$H=H_c=1.3030$ (Case 1), $H=H_c=1.27174$ (Case 2) and $H=H_c=0.1431$
(Case 3), the behavior of curves is seen through the CPs $(r_c,\pm {v_c})$.
It is shown that the fluid outflow starts from the horizon and induces by the high pressure.
The curves behaviors shown in Fig. \textbf{4} are not all physical.
For increasing radius $r$, the region must be $v>0$ (positive),
while for decreasing radius the region must be $v<0$ (negative).
The flow in the yellow, magenta, orange and purple curves
is unphysical. The fluid flow increases as $v>0$
and decreases the radius, so there is neither fluid outflow nor an accretion.
Only the red curves display the supersonic accretion in the region $v>v_c$
and subsonic accretion in the region $v<v_c$.
It is also noted here that the fluid elements are closer to
Case \textbf{1} and Case \textbf{2} instead of Case \textbf{3}, respectively. Consequently, it is viewed
that the CPs are closer for Case \textbf{1} and Case \textbf{2} as compare to Case \textbf{3},
respectively. It is revealed here that the fluid experiences the
particle emission or fluid outflow for $v>0$ while fluid accretes
for $v<0$. Specially, in Case \textbf{3} (Schwarzschild case), only the orange curve represents the
unphysical while the other four color curves show the physical behavior
which is clear as compare to Case \textbf{1} and Case \textbf{2}.
Further, Fig. \textbf{4} shows the following four
key points. \begin{itemize}
\item We notice, the subsonic/supersonic accretion occurs in the
ranges $-v_c<v<0$ and $-1<v<-v_c$, whereas supersonic/subsonic fluid outflows
for $v_c<v<1$ and $0<v<v_c$, respectively.
\item The emission of particles for $v>v_c$ and thus purely
supersonic accretion for $v<-v_c$.
\item The subsonic outflow followed by the subsonic accretion with
$v_c>v>-v_c$.
\item The upper plot shows the supersonic outflow followed by subsonic motion,
while the lower plot shows the subsonic accretion followed by supersonic accretion.
\end{itemize}
\begin{figure}
\begin{center}
\includegraphics[width=80mm]{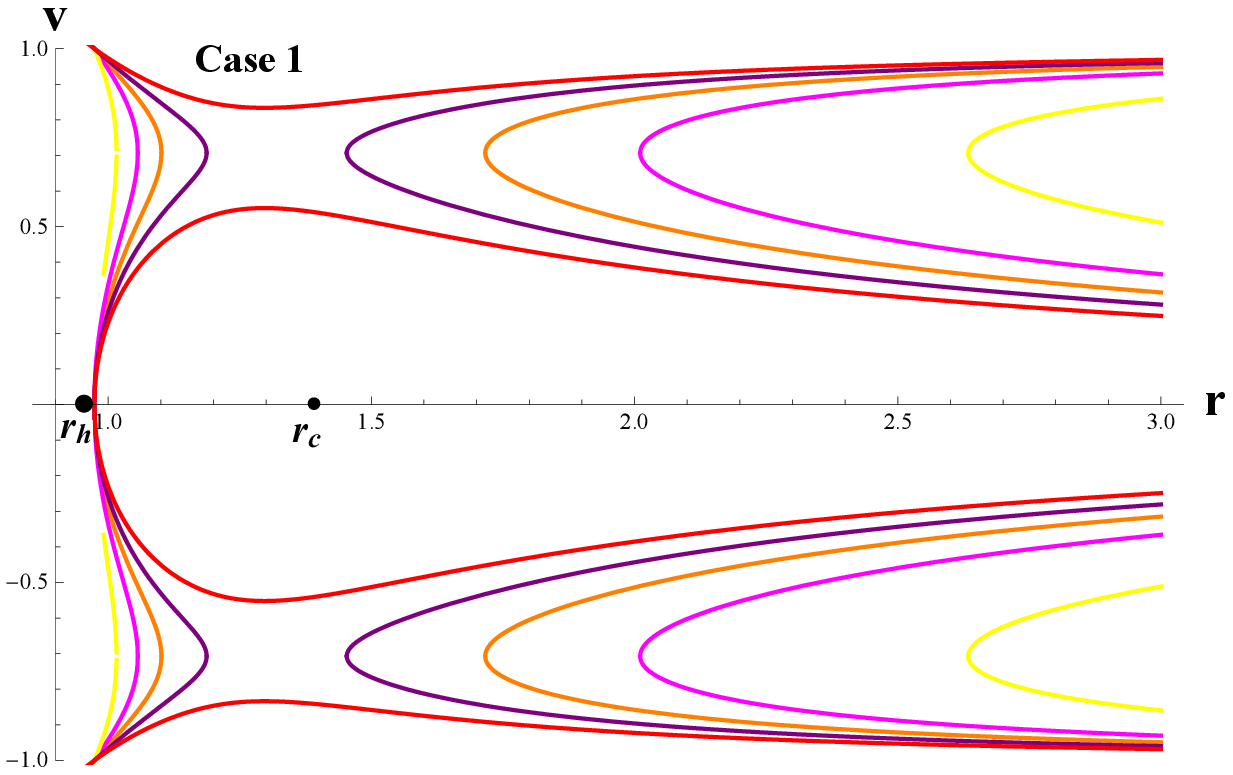}
\includegraphics[width=80mm]{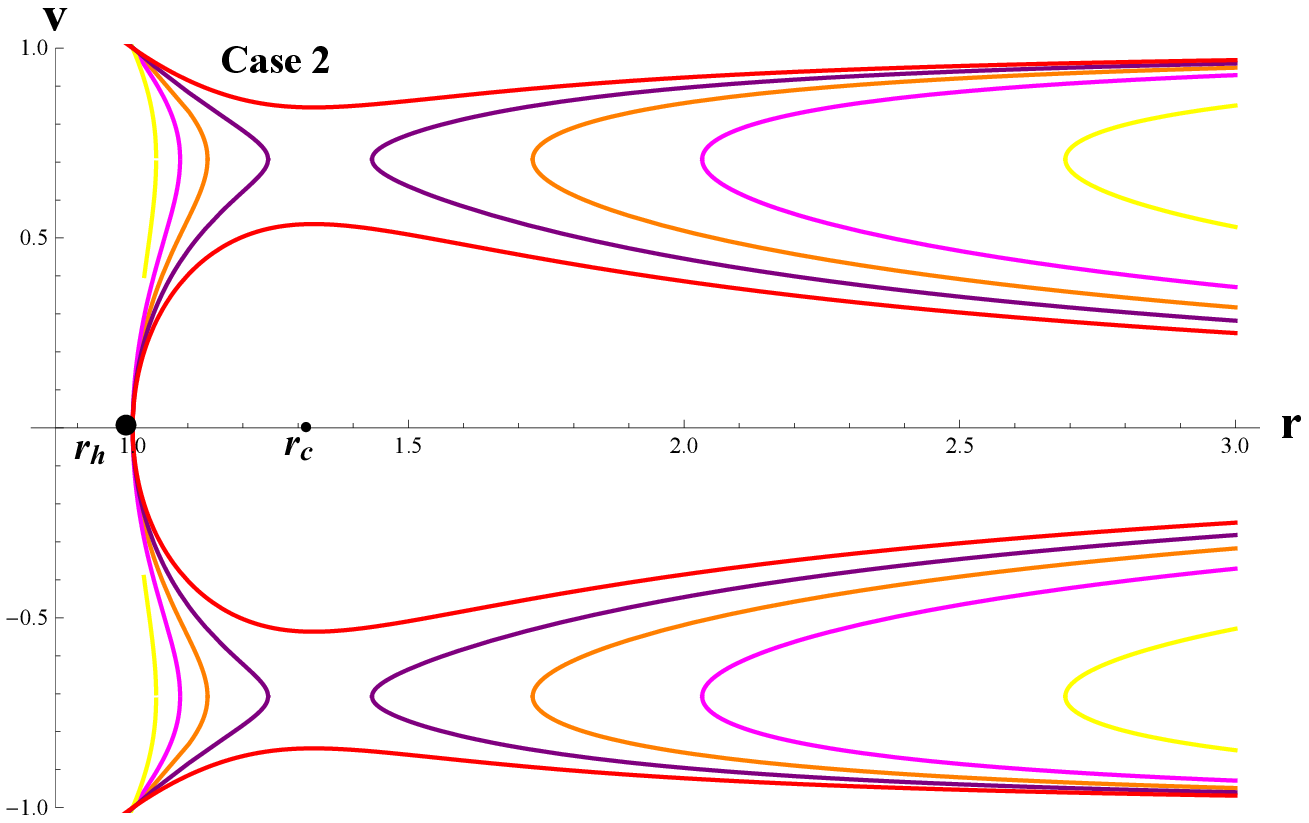}
\includegraphics[width=80mm]{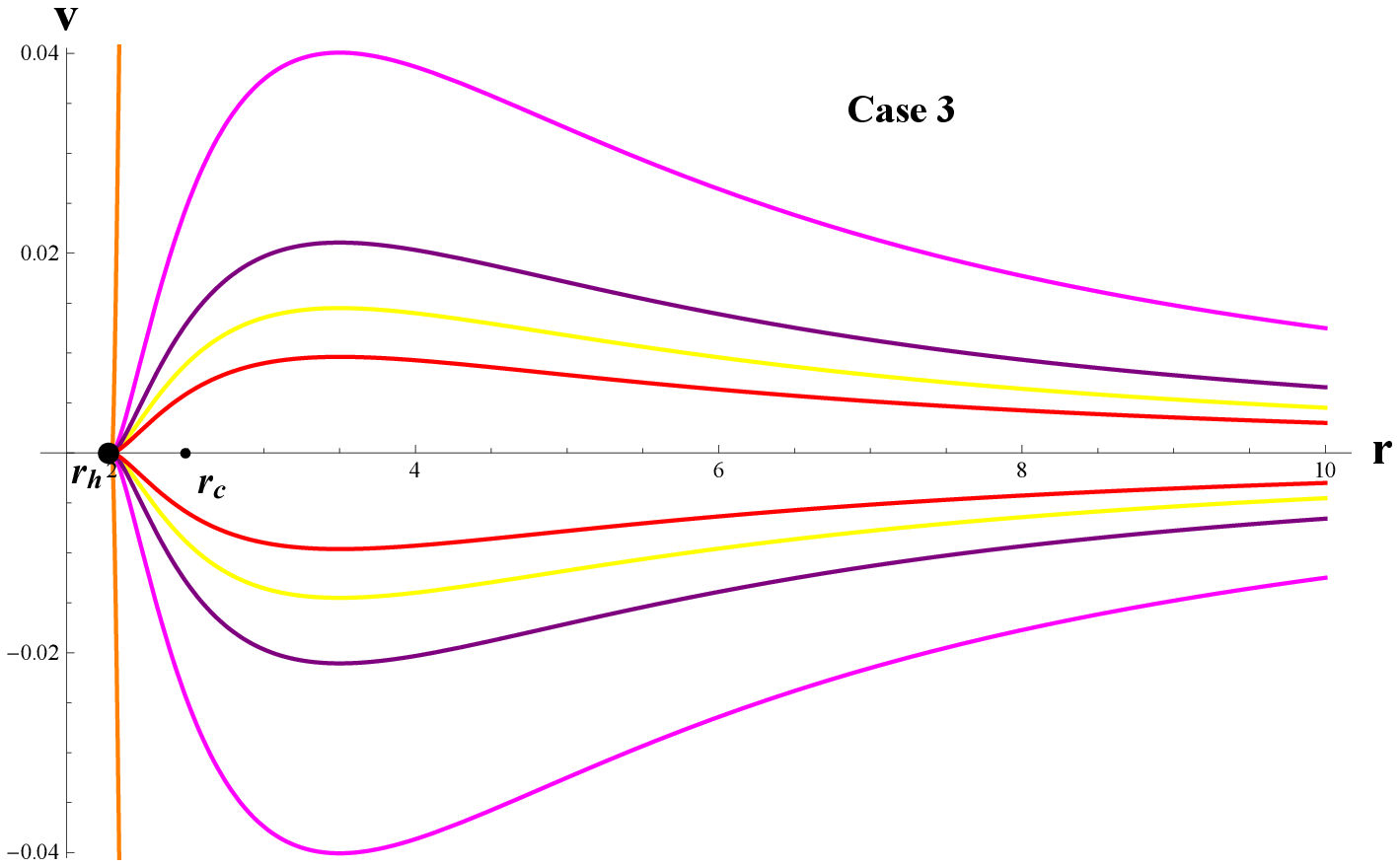}
\caption{For the accretion flow, the left panel (conformal gravity BH)
displays the behavior of Eq.($39$) with conformal parameters $\gamma=0.1$, $k=-1$, $\beta=1$.
The critical parameters involve $r_c\approx1.29701$, $v_c=0.70716$, $H_c\approx1.30306$.
Right panel (Schwarzschild de-Sitter BH) displays the behavior of ($39$)
with Schwarzschild de-Sitter BH parameters $\gamma=0$, $k=-1$, $\beta=1$.
The values of the critical parameters are $r_c\approx1.32827$, $v_c=0.70716$, $H_c\approx1.27174$.
Bottom panel (Schwarzschild BH) displays the behavior of Eq.($39$) with Schwarzschild BH parameters
$\gamma=0$, $k=0$, $\beta=1$.  The values of the critical parameters are $r_c=2.5$, $v_c=0.70716$, $H_c\approx0.143108$.
The representation of colors in $H=H_c\rightarrow$ orange, $H>H_c\rightarrow$ magenta and yellow,
$H<H_c\rightarrow$ purple and red (figure color online).}
\end{center}
\end{figure}
Consequently, we observe that the starting point of the fluid
outflow is at horizon due to its very high pressure which influences
to divergence and as a result, the fluid with its own pressure
flows back to spatial infinity \cite{58}. We also
observe from Fig. \textbf{4}, the supersonic accretion (fluid outflow) followed by
subsonic accretion (fluid inflow) stops inside the horizon and it does not give support for the claim
that ``the flow must be supersonic at the horizon" \cite{58x}. It means that for conformal gravity BH,
Schwarzschild de-Sitter BH and Schwarzschild BH the flow of the fluid is neither supersonic nor transonic near the
horizon \cite{59,60}. These results agree with fine tunning and instability issues in dynamical systems.
The stability issue is connected to the nature of the saddle points (CPs ($r_c,v_c$)
and ($r_c,-v_c$)) of the Hamiltonian system. The analysis of
stability could be done by using Lyapunov's theorem or linearization
of dynamical system \cite{61,62,63} and their variations
\cite{64}. Another stability issue is the outflow of the fluid
starts in the surrounding of horizon under the effect of pressure
divergent. This outflow is unstable because it follows a subsonic
path passing through the saddle point ($r_c, v_c$) and becomes
supersonic with speed approaches the speed of the light. The point
($r=r_h, v=0$) can be observed as attractor as well as repeller
where solution curves converge and diverge in the
cosmological point of view \cite{58x,64}.
\begin{figure}
\begin{center}
\includegraphics[width=80mm]{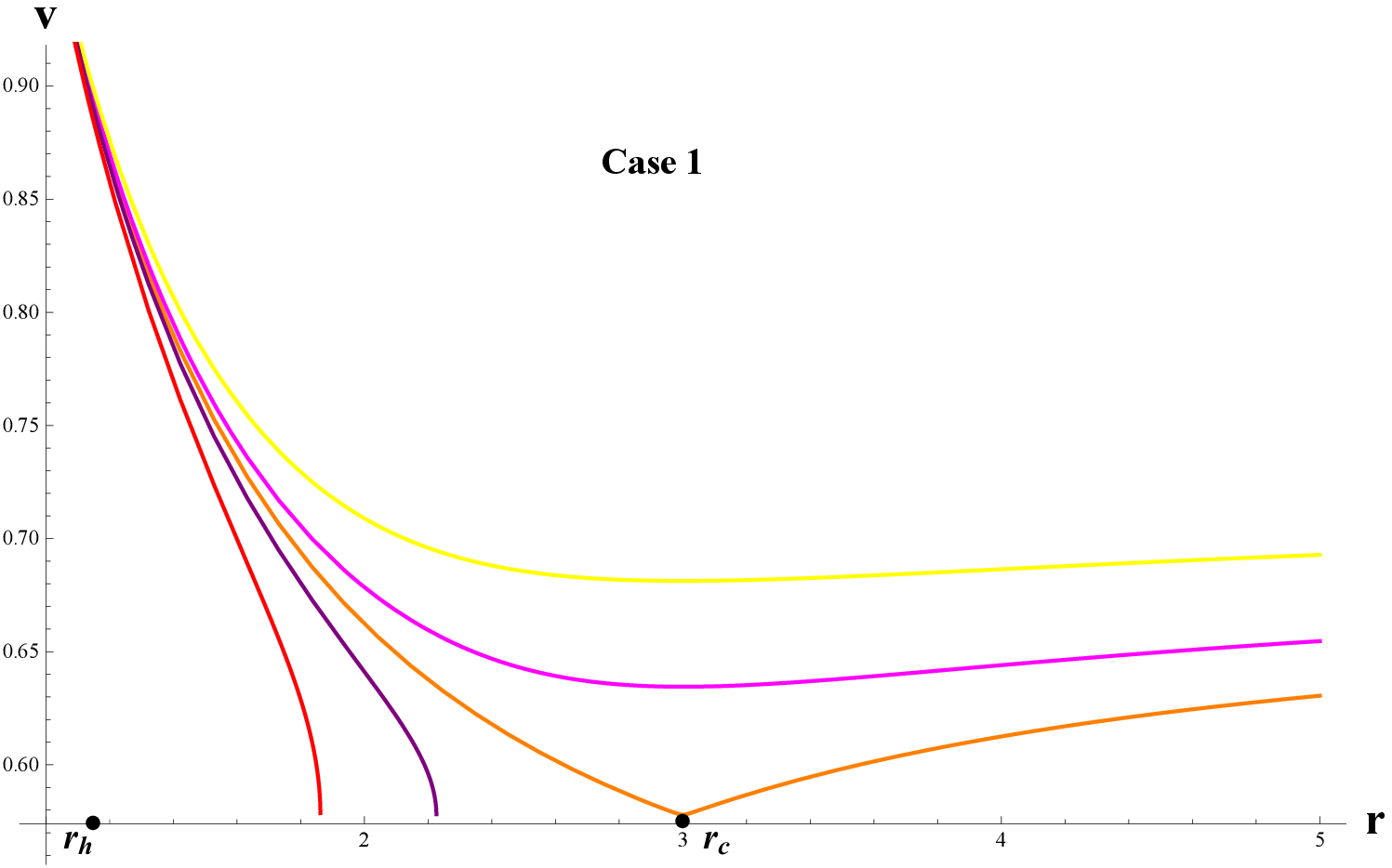}
\includegraphics[width=80mm]{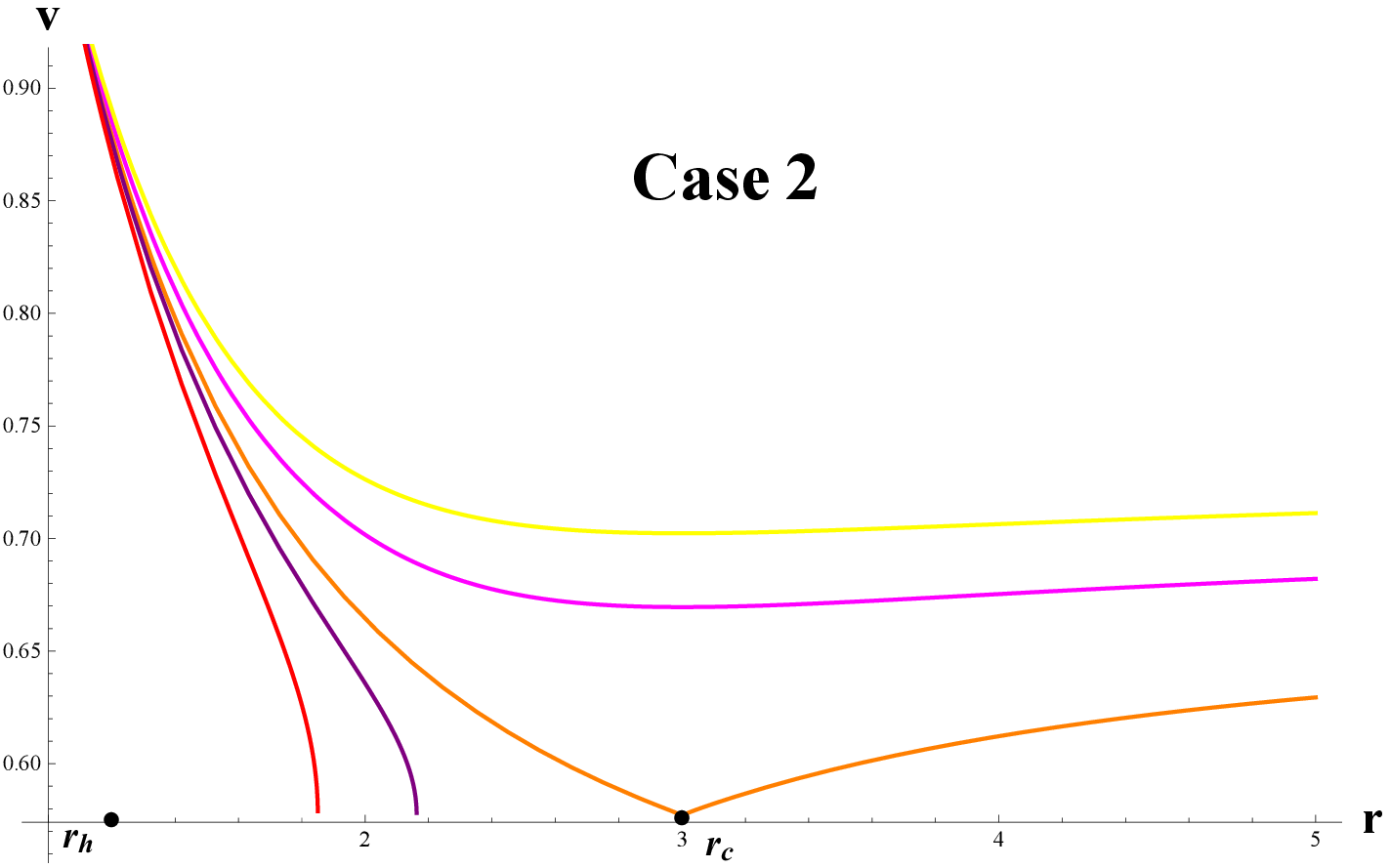}
\includegraphics[width=80mm]{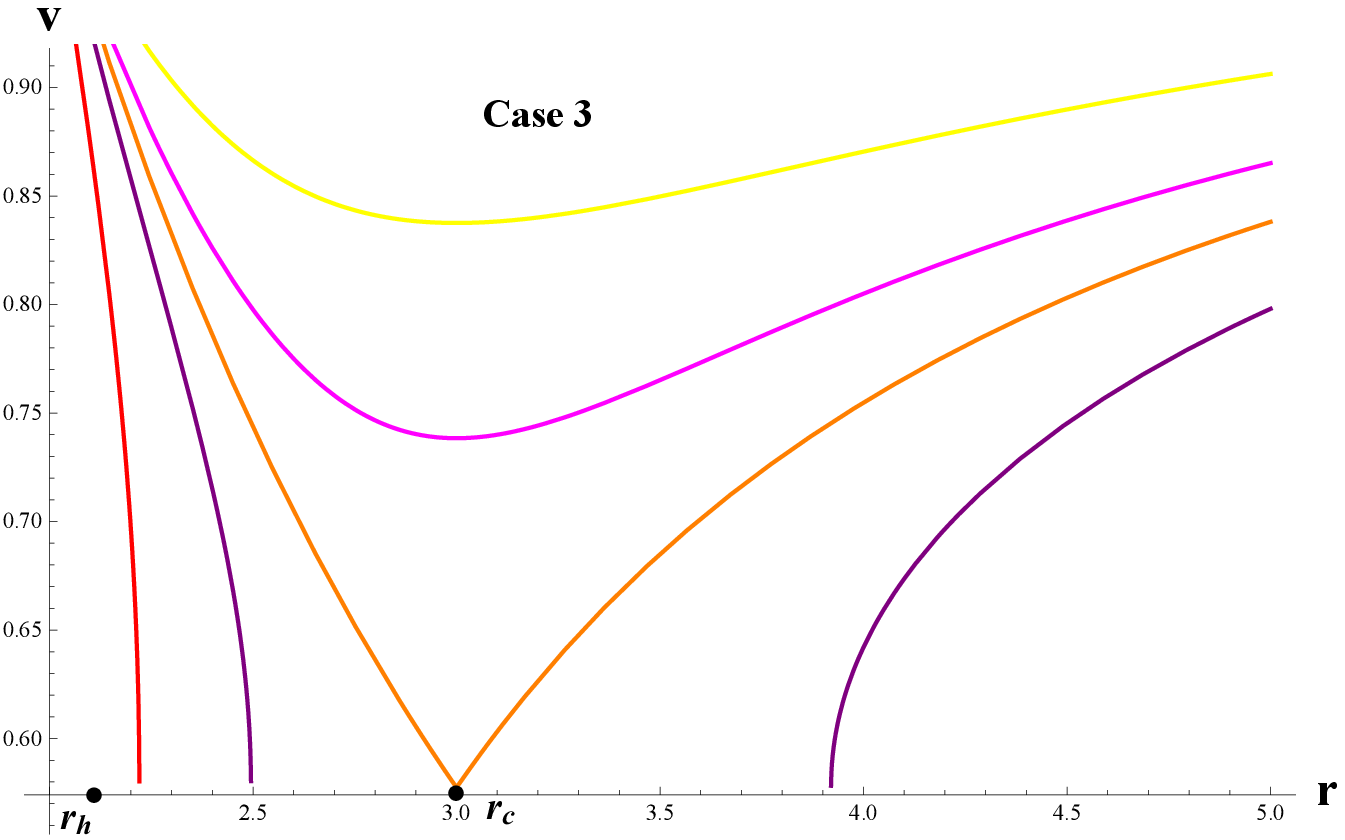}
\caption{In the physical structure of  accretion, left panel (conformal gravity BH)
displays the behavior of ($42$) with conformal parameters $\gamma=0.1$, $k=-1$, $\beta=1$.
The critical parameters involve $r_c\approx3$, $v_c=0.5773$, $H_c\approx1.95007$.
Right panel (Schwarzschild de-Sitter BH) displays the behavior of ($42$)
with Schwarzschild de-Sitter BH parameters $\gamma=0$, $k=-1$, $\beta=1$.
The critical parameters involve $r_c\approx3$, $v_c=0.5773$, $H_c\approx1.93626$.
Bottom panel (Schwarzschild BH) displays the behavior of ($42$) with Schwarzschild BH parameters
$\gamma=0$, $k=0$, $\beta=1$.  The critical parameters involve $r_c=3$, $v_c=0.5773$, $H_c\approx0.20998$.
The representation of colors in $H=H_c\rightarrow$ orange, $H>H_c\rightarrow$ magenta and yellow,
$H<H_c\rightarrow$ purple and red (figure color online).}
\end{center}
\end{figure}
\begin{figure}
\begin{center}
\includegraphics[width=80mm]{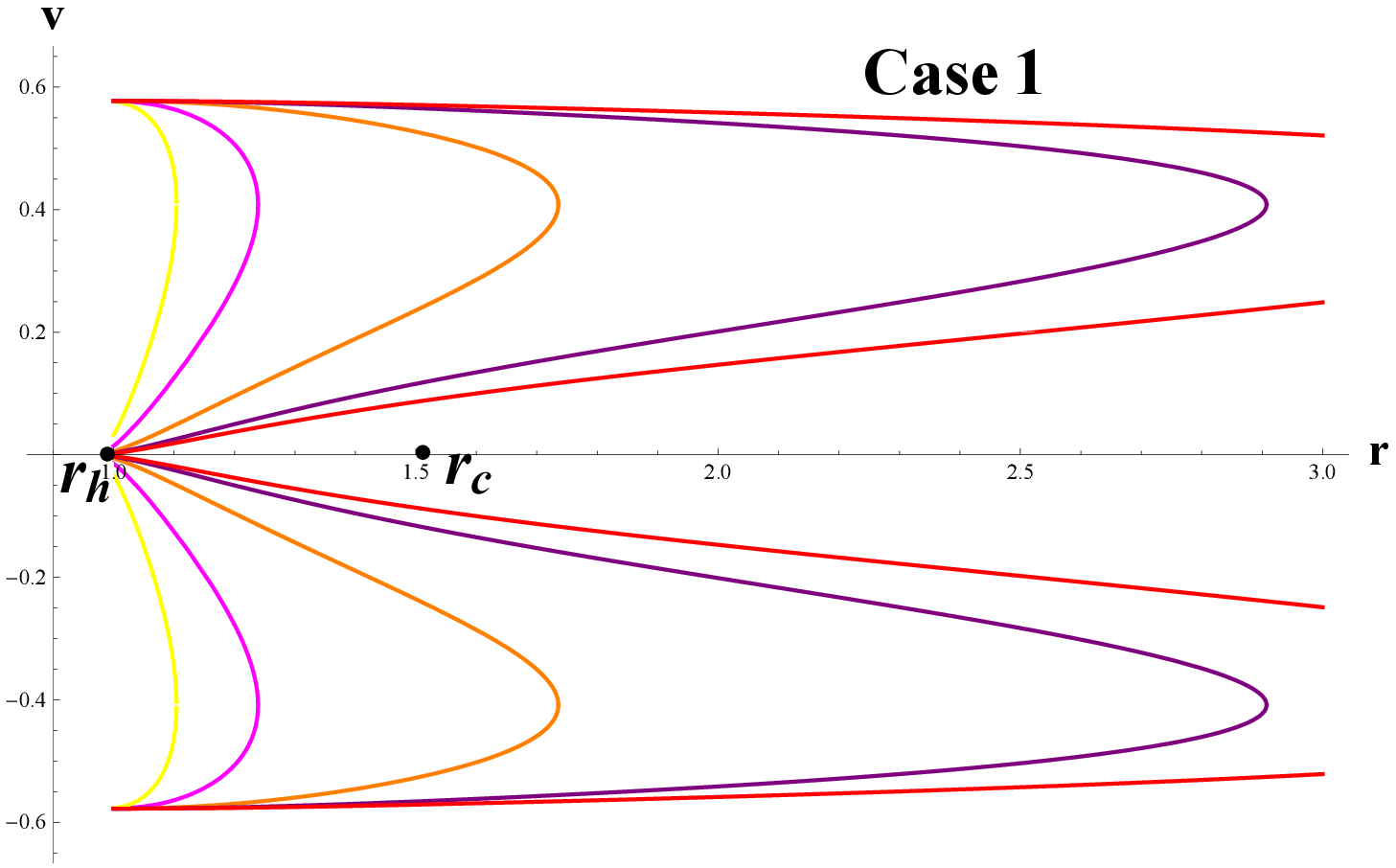}
\includegraphics[width=80mm]{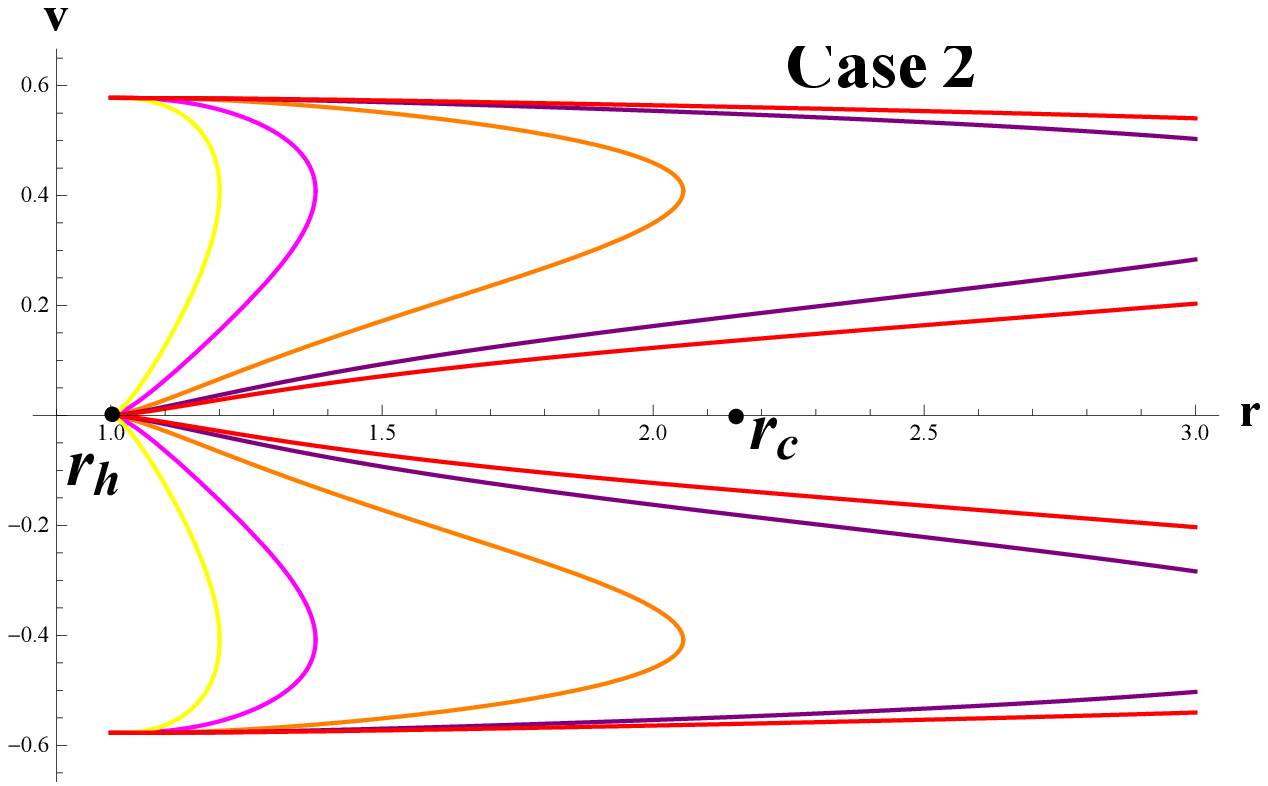}
\includegraphics[width=80mm]{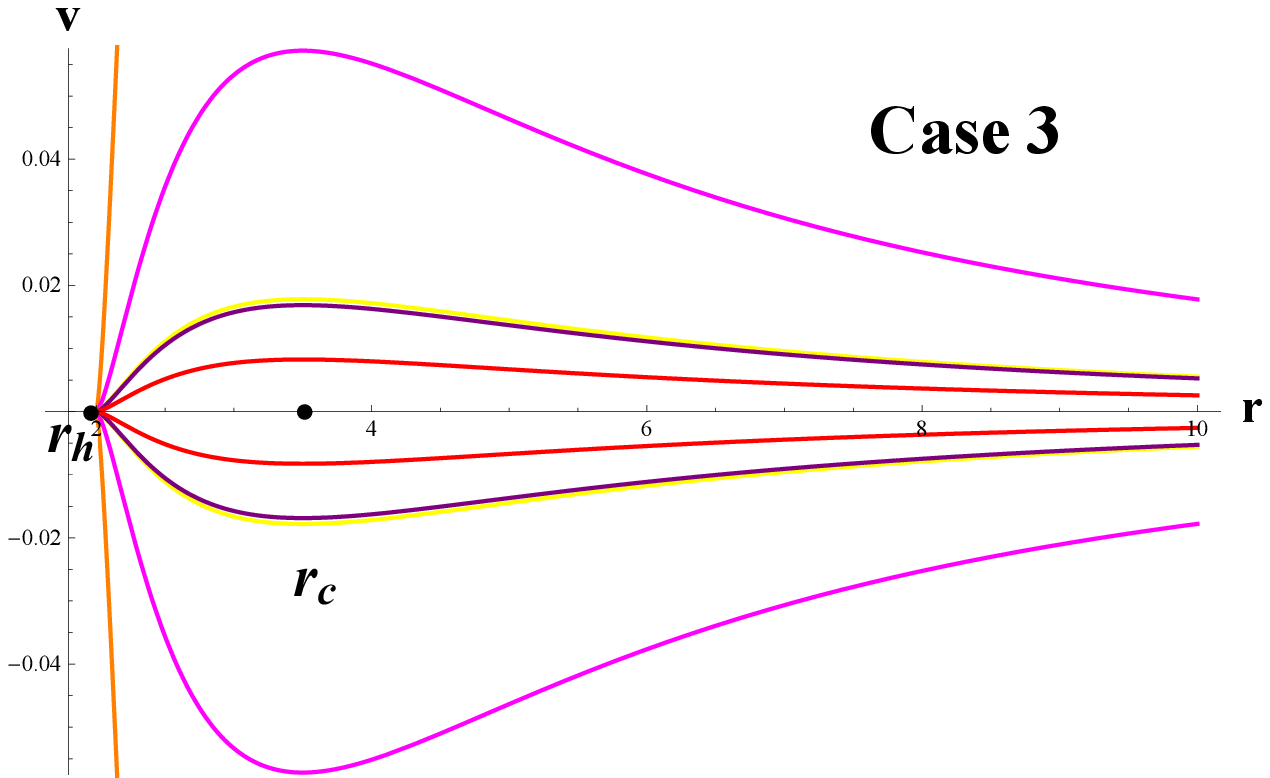}
\caption{In the physical structure of  accretion, left panel (conformal gravity BH)
displays the behavior of ($44$) with conformal parameters $\gamma=0.1$, $k=-1$, $\beta=1$.
The critical parameters involve $r_c\approx1.54267$, $v_c=0.5$, $H_c\approx2.38717$.
Right panel (Schwarzschild de-Sitter BH) displays the behavior of ($44$)
with Schwarzschild de-Sitter BH parameters $\gamma=0$, $k=-1$, $\beta=1$.
The critical parameters involve $r_c\approx2.25826$, $v_c=0.5$, $H_c\approx2.6812$.
Bottom panel (Schwarzschild BH) displays the behavior of ($44$) with Schwarzschild BH parameters
$\gamma=0$, $k=0$, $\beta=1$.  The critical parameters are taken as $r_c=3.5$, $v_c=0.5$, $H_c\approx0.26556$.
The representation of colors in $H=H_c\rightarrow$ orange, $H>H_c\rightarrow$ magenta and yellow,
$H<H_c\rightarrow$ purple and red (figure color online).}
\end{center}
\end{figure}
  \item \textbf{Radiation fluids ($\omega=\frac{1}{3}$):}
\newline
The rule played by the different parameters for the velocity profile
$v$ versus radius $r$ is important in Fig. \textbf{5}. The settlement of curves
corresponding to $H=H_c$ is colored Orange. The Magenta and
Yellow curves relate to $H<H_c$ and the Purple and
Red plots to $H>H_c$. In Fig. \textbf{5}, we
see the supersonic outflows of the fluid in the range
$v_c<v<1$. In this flow, we observe that the orange curve exactly passes
through the critical point ($r_c=3.0$) as compared to other curves.
The critical points are very close in Case \textbf{1} and Case \textbf{2} than Case \textbf{3}.
So, the orange, magenta and yellow curves are purely supersonic outflows for $(v>v_c)$ and
these curves pass through the critical speed Fig. \textbf{6} in Schwarzschild BH.
One can see the similar behavior for conformal gravity BH and Schwarzschild de-Sitter BH.
The vertical lines which are closer to the horizon are unphysical for $H>H_c$.
\item \textbf{Sub-relativistic fluids ($\omega=\frac{1}{4}$):}
\newline
We have analyzed the accretion of the sub-relativistic fluid $\omega=\frac{1}{4}$
for the conformal gravity BH. We have plotted the sub-relativistic fluid
motion versus the radius in Fig. \textbf{6}.
This figure shows that all the solution curves are not passing through the
critical velocity, which confirms to the new solution in Schwarzschild BH.
Since, the critical velocity is located at $v_c\approx0.5$ but the maximum speed in case of Schwarzschild BH approaches to $v=0.06$. So, there is no accretion flow around
Schwarzschild BH for sub-relativistic fluid.
We have observed the supersonic accretion at $v>v_c$ followed by subsonic accretion at
$0<v<v_c$ which stop at the horizon for conformal gravity BH and Schwarzschild de-Sitter BH. Furthermore, we have
the supersonic accretion with $v<-v_c$ followed by the
subsonic accretion at $0<\nu<-\nu_c$.
\end{enumerate}

\section{Polytropic Fluids Accretion}
The polytropic equation of state \cite{42, 43, 44} is
\begin{equation}
p=G(n)=\Gamma n^\alpha,\label{53}
\end{equation}
where $\Gamma$ and $\alpha$ are constants. One can consider the general
constraint $\alpha>1$ for an ordinary matter.
The specific enthalpy can be define \cite{44} as
\begin{equation}
h=m+\frac{\Gamma\alpha n^{\alpha-1}}{\alpha-1}.\label{54}
\end{equation}
Three dimensional sound speed with the help of enthalpy is given by
\begin{equation}
a^2=\frac{(\alpha-1)U}{m(\alpha-1)+U},\label{55}
\end{equation}
where $U=\gamma\alpha n^{\alpha-1}$.
Another useful result can be obtained with the help of speed of sound, which is given by
\begin{equation}
h=m\frac{\alpha-1}{\alpha-1-a^2},\label{56}
\end{equation}
and therefore
\begin{equation}
h=m\left(1+X\left(\frac{1-v^2}{r^4A(r)v^2}\right)^{(\alpha-1)/2}\right),\label{57}
\end{equation}
where
\begin{equation}
X=\frac{\Gamma \alpha n_c ^{\alpha-1}}{m(\alpha-1)}\left(\frac{r^5_c A'(r_c)}{4}\right)^{(\frac{\alpha-1}{2})}=constant>0,\label{58}
\end{equation}
and $X>0$ is a constant. From the above result, it is clear that
the constant $X$ depends on the BH parameters and also on the test fluids.
The final form of Hamiltonian system can be obtained by putting Eq. (\ref{57}) into (\ref{31}), which is given by
\begin{equation}
H=\frac{A(r)}{1-v^2}\left[1+X\left(\frac{1-v^2}{r^4A(r)v^2}\right)^{(\alpha-1)/2}\right]^2,\label{59}
\end{equation}

1. Hamiltonian for conformal gravity BH
\begin{eqnarray}\label{60}
H=&&\frac{\Big(1-\frac{\beta(2-3\beta\gamma)}{r}-3\beta\gamma+\gamma r-kr^2\Big)}{1-v^2}\Big[1
\\\nonumber&&+X\Big(\frac{1-v^2}{r^4v^2\Big(1-\frac{\beta(2-3\beta\gamma)}{r}-3\beta\gamma+\gamma r-kr^2\Big)}\Big)^{(\alpha-1)/2}\Big]^2.
\end{eqnarray}

2. Hamiltonian for Schwarzschild de-Sitter BH
\begin{eqnarray}\label{61}
H=&&\frac{\Big(1-\frac{2\beta}{r}-kr^2\Big)}{1-v^2}\Big[1
+X\Big(\frac{1-v^2}{r^4v^2(1-\frac{2\beta}{r}-kr^2)}\Big)^{(\alpha-1)/2}\Big]^2.
\end{eqnarray}

3. Hamiltonian for Schwarzschild BH
\begin{eqnarray}\label{62}
H=&&\frac{1-\frac{2\beta}{r}}{1-v^2}\Big[1
+X\Big(\frac{1-v^2}{r^4v^2(1-\frac{2\beta}{r})}\Big)^{(\alpha-1)/2}\Big]^2.
\end{eqnarray}
It is analyzed from the Hamiltonian results that $\frac{dA(r)}{dr}>0$ for all $r$.

Adopting the technique in \cite{42, 43, 44}, one can get the following relation
\begin{equation}\label{63}
(\alpha-1-v^2_c)\left(\frac{1-v^2_c}{r^4_cA(r_c)v^2_c}\right)^{\frac{\alpha-1}{2}}=\frac{n_c}{2X}\left(r^5_cA'(r_c)\right)^{\frac{1}{2}v^2_c},
\end{equation}
\begin{equation}\label{64}
v^2_c = \frac{r_cA'_{rc}}{r_cA'_{r_c}+4A(r_c)}.
\end{equation}

\begin{figure}
\begin{center}
\includegraphics[width=80mm]{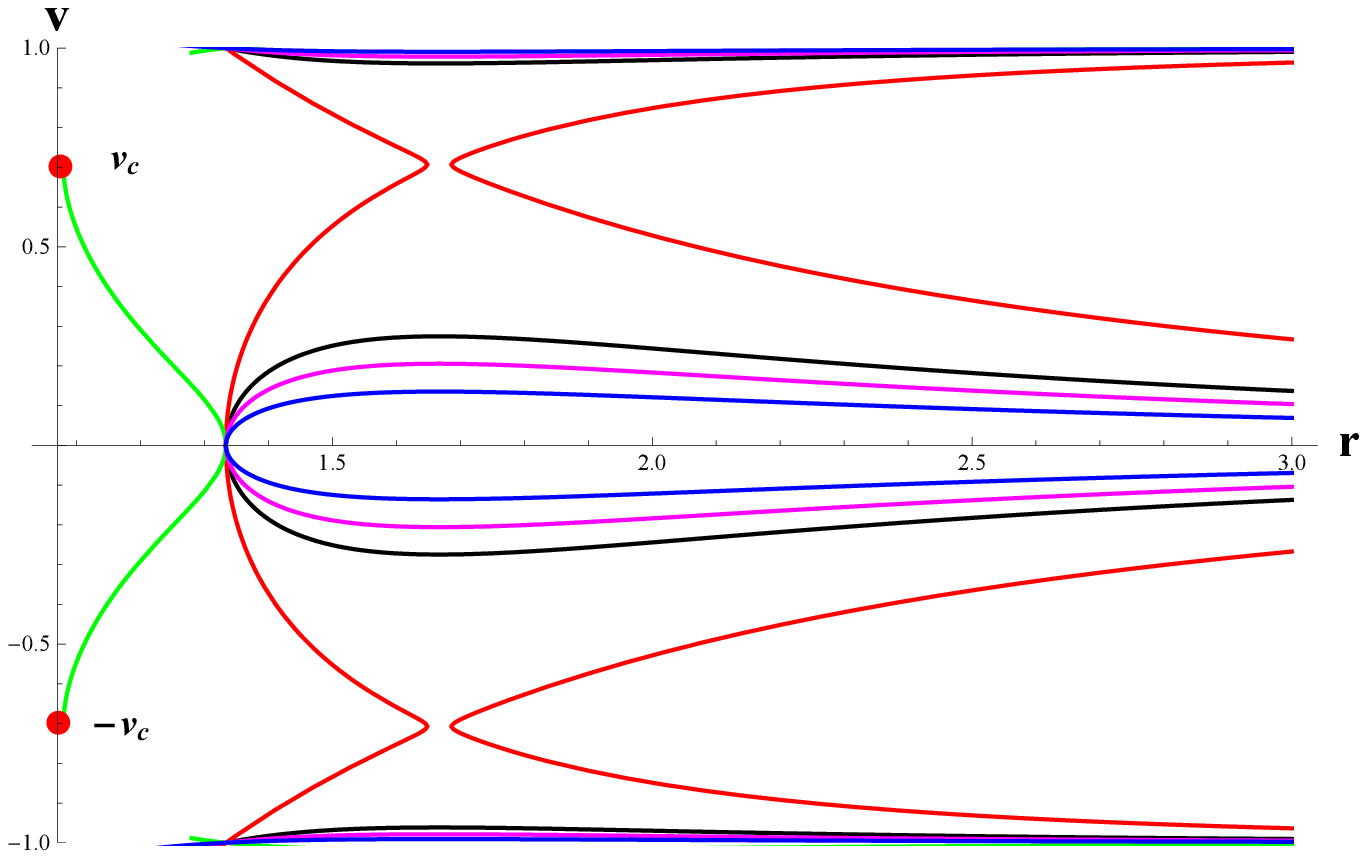}
\includegraphics[width=80mm]{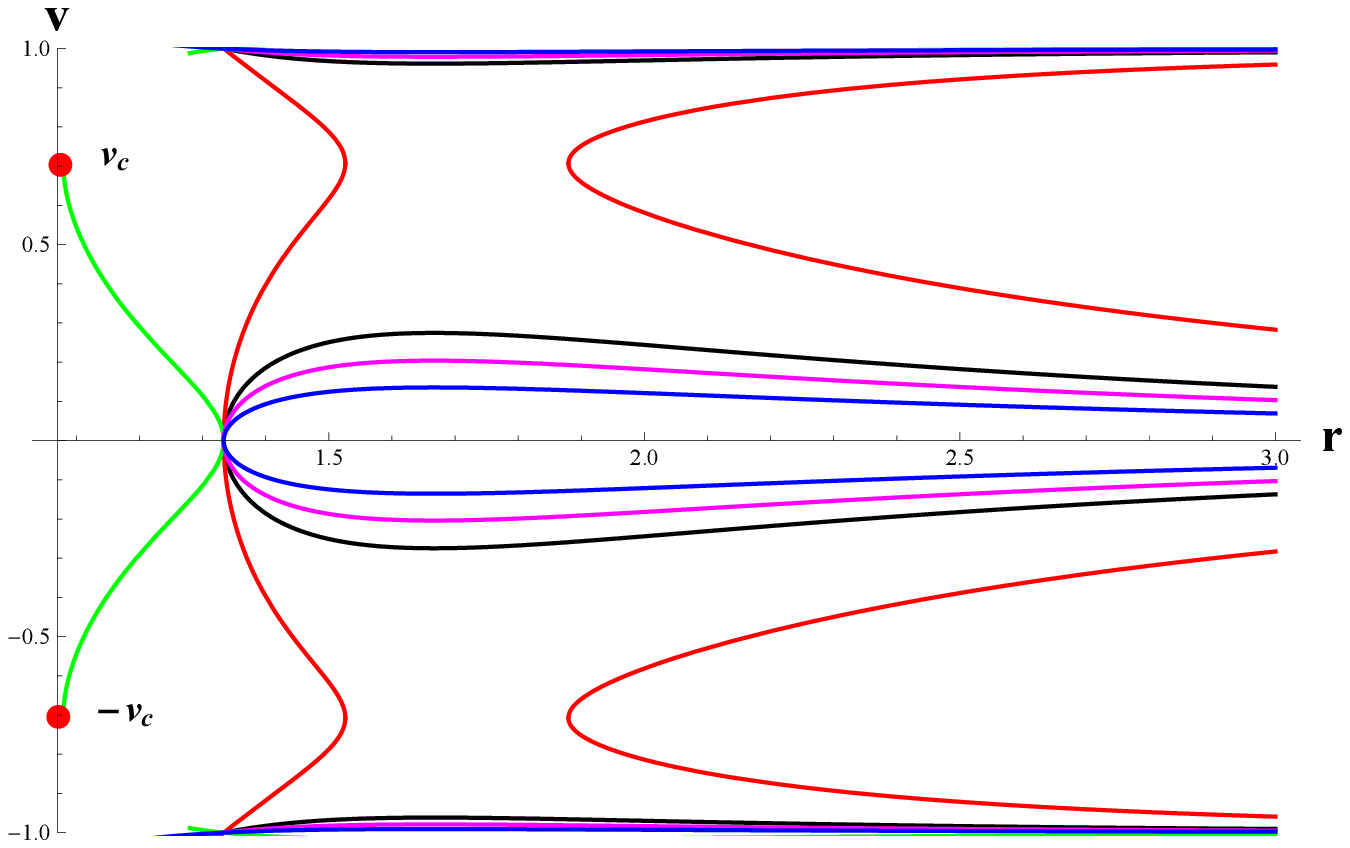}
\includegraphics[width=80mm]{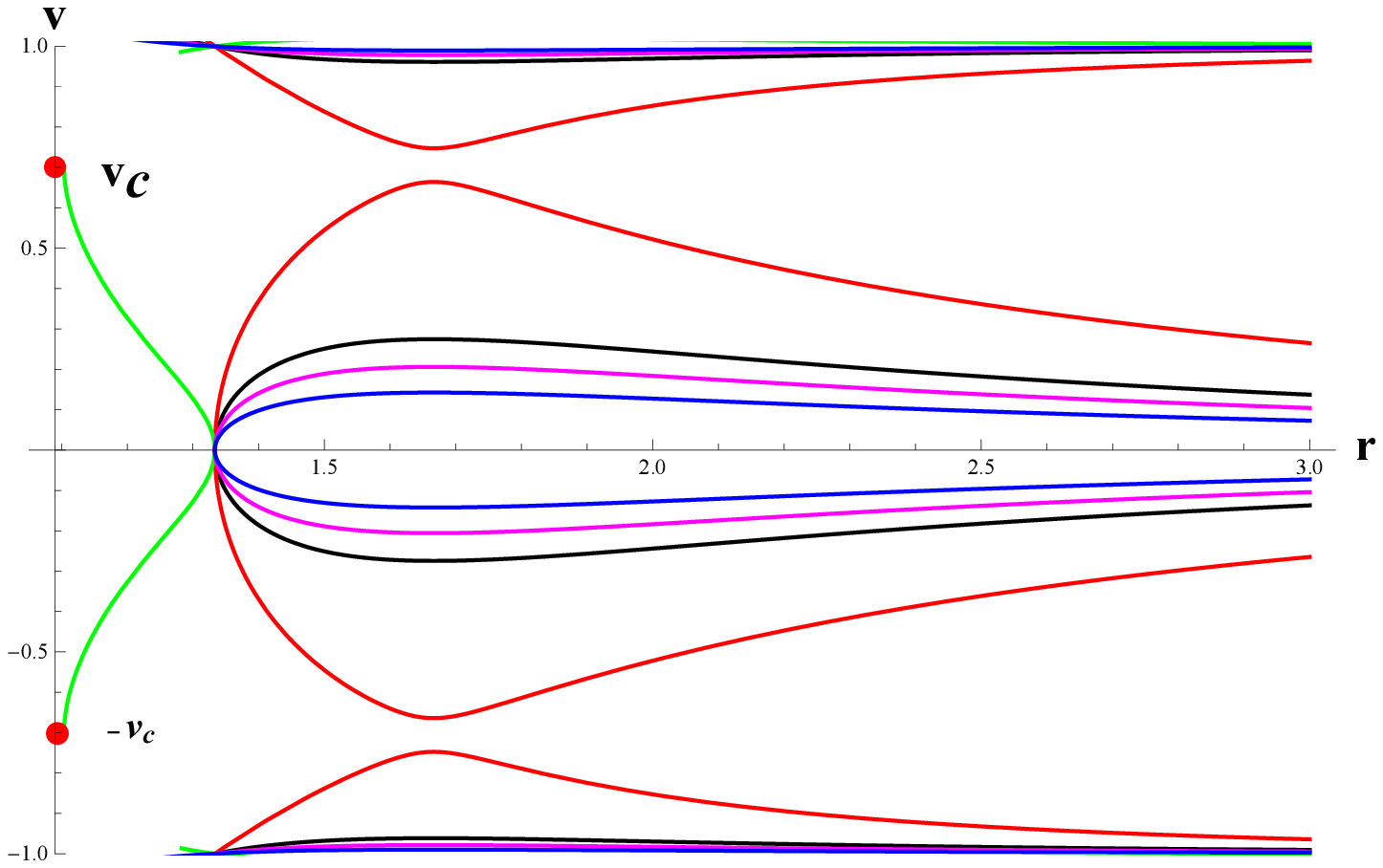}
\caption{For the polytropic fluid accretion, left panel (conformal gravity BH)
displays the behavior of (\ref{60}). Right panel (Schwarzschild de-Sitter BH)
displays the behavior of (\ref{61}). Bottom panel (Schwarzschild BH)
displays the behavior of (\ref{62}) (figure color online).}
\end{center}
\end{figure}

Figure \textbf{7} represents the contour plots for conformal BH (left panel),
Schwarzschild de-Sitter BH (right panel) and Schwarzschild BH (bottom panel)
with $n_c=0.15$, $X=5$, $\alpha=5/3$. We have presented the behavior of matter
by taking the sonic points $r_c\approx1.9855$, $v_c\approx0.56218$,
$H=H_c\simeq 4.2876$ for conformal BH, $r_c\approx3.7859$, $v_c\approx0.44216$,
$H=H_c\simeq 2.1377$ for Schwarzschild de-Sitter BH and $r_c\approx5.2865$, $v_c\approx0.24211$,
$H=H_c\simeq 1.9374$ for Schwarzschild BH. It is analyzed that the critical flow for
conformal BH, Schwarzschild de-Sitter BH and Schwarzschild BH starts from
subsonic accretion and then follows the supersonic accretion escaping the saddle
point (sonic point) and ends at the Killing horizon. The supersonic accretion
begins from the region of Killing horizon and ends at the subsonic accretion as
$r$ approaches to infinity. It has been observed that the accretion behavior of various BHs
is different at critical points of polytropic test fluids case. Also, it has been observed
that at the trajectory of conformal BH, the critical points are closer (see red curve),
for Schwarzschild de-Sitter BH, the critical points are also distant (see red curve)
and for Schwarzschild BH, the critical points are also distant (see red curve).
In all these cases, the trajectories do not pass through the saddle point (sonic point).

\section{Black Hole's Mass Accretion Rate}
The mass accretion rate of BH is an important aspect in the study of accretion, for this purpose, we have calculated the accretion rate corresponding to $A(r)$. Specially, we have observed the effects of radius
on the accretion rate.
Generally, mass accretion rate is the area times flux at the boundary of BH
and is denoted by $\dot{M}$, it evaluates the BH mass per unit time.
Here, we consider the general expression for the mass accretion rate
as $\dot{M}\mid_{rh}=4\pi r^2 T^r_t \mid_{rh}$ \cite{57}, the energy momentum tensor for perfect fluid
can be as used. Since, the dynamical system is conserved
$\Delta_\mu J^\mu=0$ and $\nabla_\nu T^{\mu\nu}=0$. Thus, due to this conserved system,
Eqs. (\ref{13}) and (\ref{17}) give
\begin{equation}
r^2u(\varrho+p)\sqrt{A(r)+(u)^2}=L_0,\label{65}
\end{equation}
where $L_0$ is the constant. Now, by taking the continuity equation (relativistic energy flux)
and the equation of state $p=p(\varrho)$, we get
\begin{equation}
\frac{d\varrho}{\varrho+p}+\frac{du}{u}+\frac{2}{r}dr=0.\label{66}
\end{equation}
After integrating, we obtain
\begin{equation}
ru \exp\left[\int^{\varrho}_{\varrho\infty}\frac{d \varrho'}{\varrho'+p(\varrho')}\right]=-L_1,\label{67}
\end{equation}
where $L_1$ is the constant of integration and $\varrho_\infty$ represents the fluid density at infinity.
Here, the minus is taken due to $u<0$.
Dividing Eq. (\ref{65}) with (\ref{67}), we get
\begin{equation}
L_3=-\frac{L_0}{L_1}=(\varrho+p)\sqrt{A(r)+(u)^2}\exp\left[-\int^{\varrho}_{\varrho\infty}\frac{d \varrho'}{\varrho'+p(\varrho')}\right],\label{68}
\end{equation}
where $L_3$ is a constant. At infinity, $L_3=\varrho_\infty +p(\varrho_\infty)=-\frac{L_0}{L_1}$,
with $L_0=(\varrho+p)u^0 u r^2=-L_1(\varrho_\infty +p(\varrho_\infty))$.
The problem is spherically symmetrically static at equatorial plane, so, the mass flux equation
$\nabla_\mu J^\mu=0$ takes the form
\begin{equation}
r^2un=L_2.\label{69}
\end{equation}
Where $L_2$ is an integration constant.
Dividing the Eq. (\ref{65}) with (\ref{69}), we get
\begin{equation}
\frac{\varrho+p}{n}\sqrt{A(r)+(u)^2}=\frac{L_0}{L_2}\equiv L_4,\label{70}
\end{equation}
where  $L_4=\frac{(\varrho_\infty+p_\infty)}{n_\infty}$.
Using Eq. (\ref{65}), the mass of BH takes the following form
\begin{equation}
\dot{M}=-4\pi r^2u(\varrho+p)\sqrt{A(r)+(u)^2}=-4\pi L_0.\label{71}
\end{equation}
Further, it takes the form
\begin{equation}
\dot{M}=4\pi L_1(\varrho_\infty+p(\varrho_\infty)).\label{72}
\end{equation}
Above equation gives the valid result for any nature of fluids.
Thus, we have
\begin{equation}
\dot{M}=4\pi L_1(\varrho+p)|_{r=r_h},\label{73}
\end{equation}
Assuming the isothermal equation of state $(p=\omega\varrho)$,
which implies that $(\varrho+p)=\varrho(1+\omega)$.
Then, using the Eq. (\ref{67}), it leads us to
\begin{equation}
\varrho=\left[-\frac{L_1}{r^2u}\right]^{1+\omega}.\label{74}
\end{equation}
By the expression of $\varrho$ in Eq. (\ref{65}) one can obtain the following general equation
\begin{equation}
(u)^2-\frac{L^2_0L^{-2(1+\omega)}_1}{(1+\omega)^2}r^{4\omega}(-u)^{2\omega}+A(r)=0.\label{75}
\end{equation}
It can be solved for fluid velocity $u$ with any value of $\omega$.
One can calculate the energy density $\varrho$ by using $u$ with $\omega$.\\
\underline{\textbf{Exact solution for ultra-stiff fluids $\omega=1$:}}\\
By assuming $\omega=1$ in Eq. (\ref{70}) and (\ref{74}), one can calculate the radial velocity and energy-density of ultra-stiff fluids, that is
\begin{equation}
u=\pm L_1^2\sqrt{\frac{A(r)}{L^2_0r^4-4L^4_1}},\label{76}
\end{equation}
also the energy density is given by
\begin{equation}
\varrho=\frac{(L^2_0r^4-4L^4_1)}{4L^2_1r^4A(r)}.\label{77}
\end{equation}
From Eqs. (\ref{73}) and (\ref{77}), the mass accretion rate of conformal gravity BH can obtained in the following form
\begin{equation}
\dot{M}=\frac{2\pi (L^2_0r^4-4L^4_1)}{L_1r^3\left[3\beta^2\gamma-2\beta+(1-3\beta\gamma)r+\gamma r^2-kr^3\right]}~~~~~~~~~~~~~~ (\rm{Case}\quad\textbf{1}).\label{78}
\end{equation}
Similarly, following the same method, we can find
the mass accretion rate for Schwarzschild de-Sitter and Schwarzschild BHs
\begin{eqnarray}
\dot{M}&=&\frac{2\pi (L^2_0r^4-4L^4_1)}{L_1r^3\left[-2\beta+r-kr^3\right]}.~~~~~~~~~~~~~~~~~~~~~~ (\rm{Case}\quad \textbf{2})\label{79}\\
\dot{M}&=&\frac{2\pi (L^2_0r^4-4L^4_1)}{L_1r^3\left[-2\beta+r\right]}.~~~~~~~~~~~~~~~~~~~~~~~~~~~~ (\rm{Case}\quad\textbf{3})\label{80}
\end{eqnarray}

\begin{figure}
\begin{center}
\includegraphics[width=80mm]{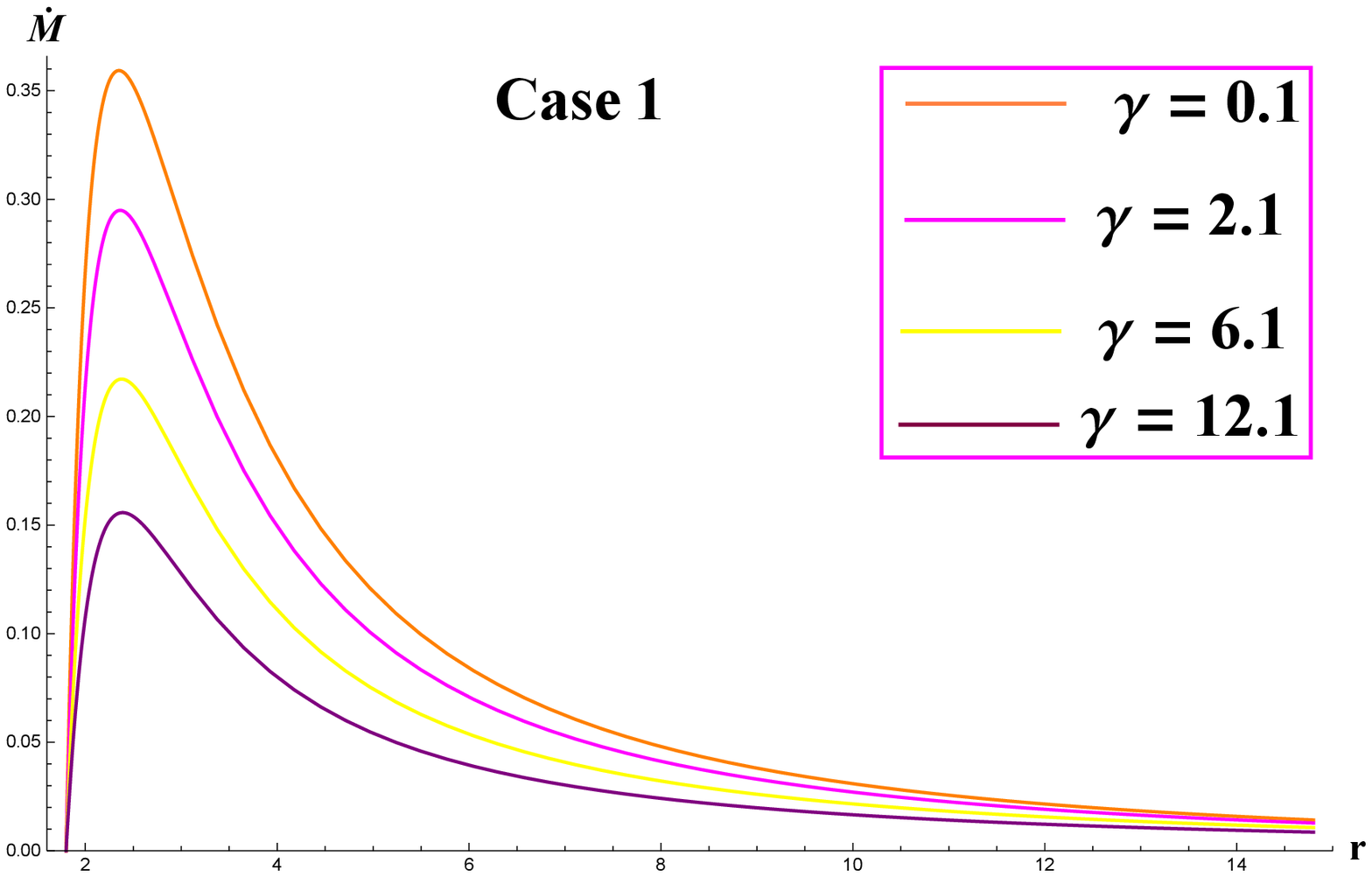}
\includegraphics[width=80mm]{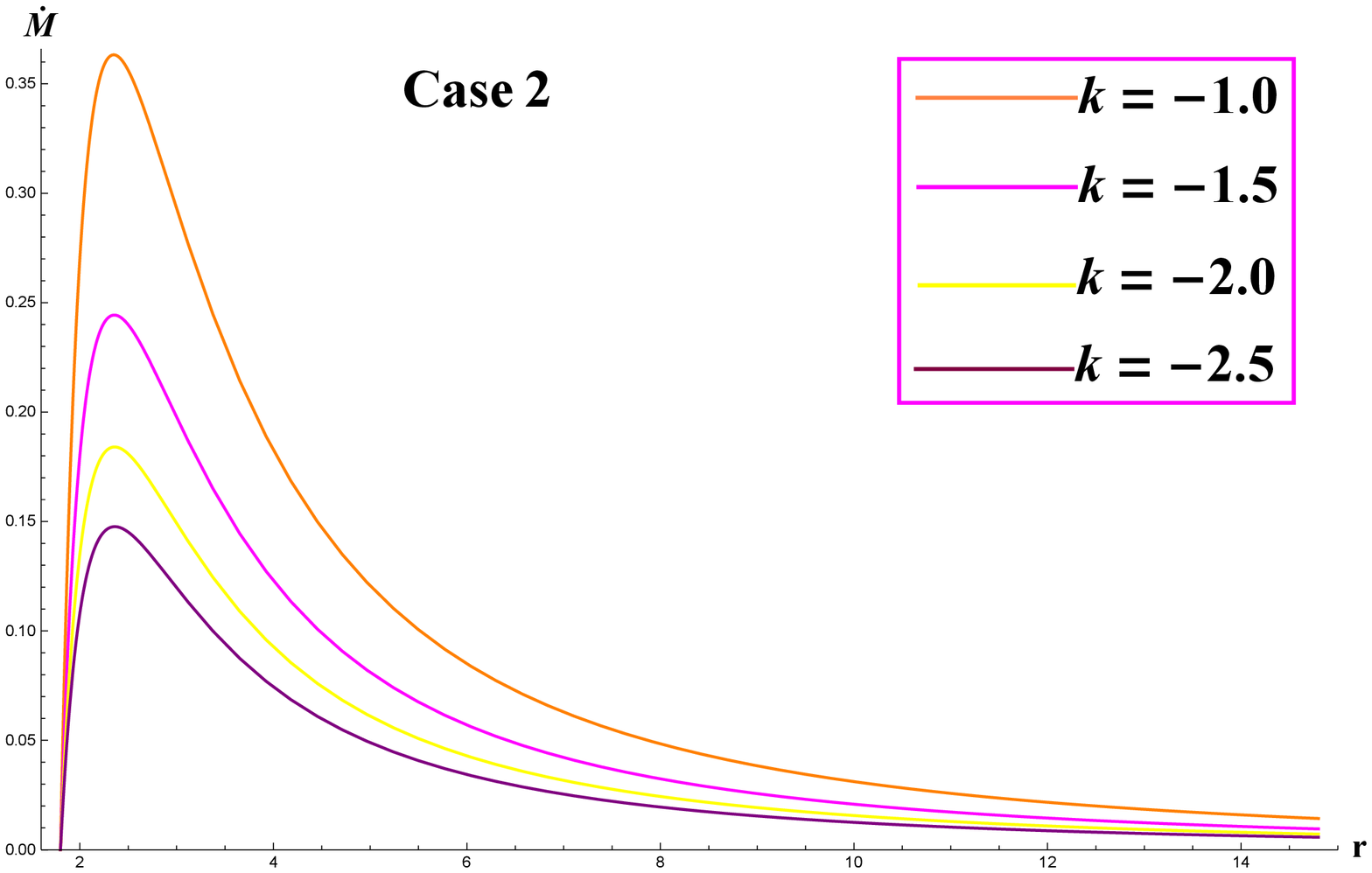}
\includegraphics[width=80mm]{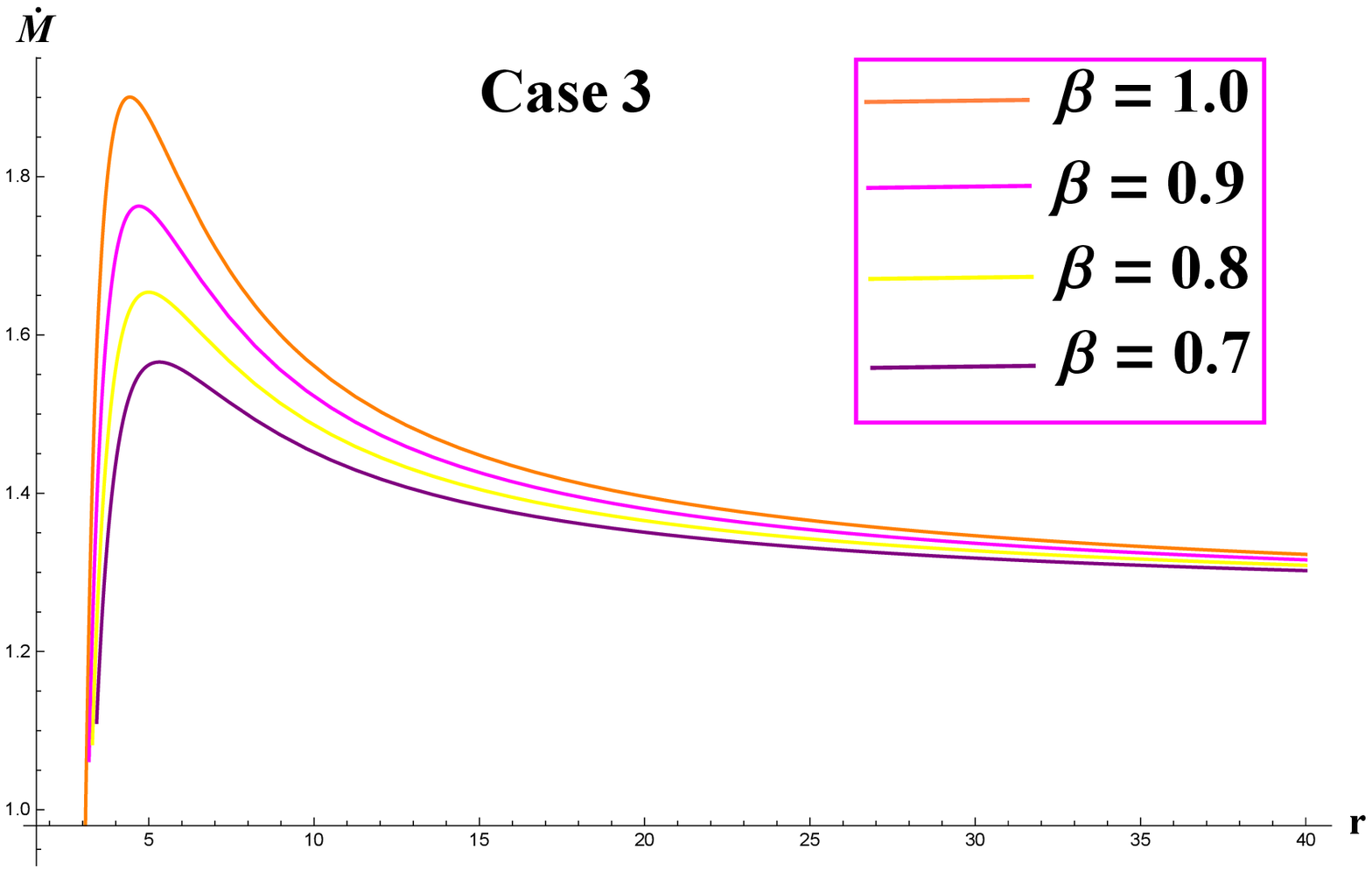}
\caption{In this figure, left panel (conformal gravity BH)
displays the behavior of Eq. (\ref{78}) with conformal parameters $\gamma=0.1$, $k=-1$, $\beta=1$.
Right panel (Schwarzschild de-Sitter BH) displays the behavior of Eq. (\ref{79})
for Schwarzschild de-Sitter BH parameters $\gamma=0$, $k=-1$, $\beta=1$.
Bottom panel (Schwarzschild BH) displays the behavior of Eq. (\ref{80}) for Schwarzschild BH
parameters $\gamma=0$, $k=0$, $\beta=1$. Other constants are taken as $L_0=0.90$, $L_1=0.5$ (figure color online).}
\end{center}
\end{figure}
In Fig. \textbf{8}, we plot the mass accretion-rate $\dot{M}$ versus the radius $r$
for aforemention BHs in ultra-stiff fluid, other parameters $\gamma$, $k$
and $\beta$ are taken as fixed. In the left panel (conformal gravity BH) the accretion rate is
increased by decreasing the parameter $\gamma$. It has been noted that the
maximum accretion rate occurs for overlapping the critical radius in the presence of different
values of $\gamma$. For the left panel the values of mass accretion rate are:
$\dot{M}=0.15, 0.235, 0.29, 0.36$ for $\gamma=12.1, 6.1, 2.1, 0.1$, we have the overlapping
radius $r\approx 1.931$, respectively. The same critical points are seen in the right panel
(Schwarzschild de-Sitter BH) in the presence of parameter $k$. The critical values are:
$\dot{M}=0.15, 0.235, 0.29, 0.36$ for $k=-2.5, -2.0, -1.5, -1.0$, we have the overlapping
radius $r\approx 1.931$, respectively. The right panel increases
the mass of Schwarzschild de-Sitter BH by increasing the cosmological constant parameter $k$.
Now, the mass of Schwarzschild BH occurs in the bottom panel where we observe the maximum
accretion rate $\dot{M}=15.5, 16.5, 17.5, 18.5$ occur for different values of parameters
\begin{itemize}
  \item $\beta=0.7$, corresponding to $r\approx 2.536$
  \item $\beta=0.8$, corresponding to $r\approx 2.936$
  \item $\beta=0.9$, corresponding to $r\approx 3.156$
  \item $\beta=1.0$, corresponding to $r\approx 3.956$
\end{itemize}
The mass accretion rate of Schwarzschild BH is increased
by increasing the mass function $(\beta)$. Hence, we conclude that the mass of
Schwarzschild BH is larger as compare to conformal gravity BH and Schwarzschild de-Sitter BH.
Also, it is concluded that the conformal parameters are critically important for
the maximum accretion rate in ultra-stiff fluids.

\underline{\textbf{Exact solution for ultra-relativistic fluids $\omega=1/2$:}}\\
By assuming $\omega=1/2$ in Eqs. (\ref{70}) and (\ref{74}), we calculate the radial velocity and the
energy-density of the radiation fluids, that is
\begin{eqnarray}
u&=&\frac{2r^2L^2_0+\sqrt{4r^2L^4_0-81A(r)L^6_1}}{9L^3_1}.\label{81}\\
\varrho&=&27\left(\frac{L^4_1}{r^2(2r^2L^2_0+\sqrt{4r^2L^4_0-81A(r)L^6_1})}\right)^{3/2}.\label{82}\\
\dot{M}&=&216\pi L_1\left(\frac{L^4_1}{r^2(2r^2L^2_0+\sqrt{4r^2L^4_0-81A(r)L^6_1})}\right)^{3/2}.\label{83}
\end{eqnarray}

 1. The mass accretion rate of conformal gravity BH is

\begin{eqnarray}\label{84}
\dot{M}=&&216\pi L_1
\\\nonumber&&\times\left(\frac{L^4_1}{r^2(2r^2L^2_0+\sqrt{4r^2L^4_0-81(1-\frac{\beta(2-3\beta\gamma)}{r}-3\beta\gamma+\gamma r-kr^2)L^6_1})}\right)^{3/2}.
\end{eqnarray}
 2. The mass accretion rate of Schwarzschild de-Sitter BH is

\begin{eqnarray}\label{85}
\dot{M}=&&216\pi L_1
\\\nonumber&&\times\left(\frac{L^4_1}{r^2(2r^2L^2_0+\sqrt{4r^2L^4_0-81(1-\frac{2\beta}{r}-kr^2)L^6_1})}\right)^{3/2}.
\end{eqnarray}

 3. The mass accretion rate of Schwarzschild BH is

\begin{eqnarray}\label{86}
\dot{M}=&&216\pi L_1
\\\nonumber&&\times\left(\frac{L^4_1}{r^2(2r^2L^2_0+\sqrt{4r^2L^4_0-81(1-\frac{2\beta}{r})L^6_1})}\right)^{3/2}.
\end{eqnarray}

In Fig. \textbf{9}, we plot the mass accretion-rate ($\dot{M}$) versus the radius ($r$)
for aforemention BHs in ultra-relativistic fluid. The left panel (conformal gravity BH) shows that the accretion rate is
decreasing for larger value of $r$, that is $\dot{M}=7000$ for $r=0.8$ and Killing
horizon is at $r_{KH}\approx0.8$ whereas the universal horizon is at $r_{UH}\approx0.1$.
The accretion rate is increasing for smaller values of the radius, that is $\dot{M}>8000$ for $r\simeq0.65$
and the Killing horizon is at $r_{KH}\approx0.65$ whereas the universal horizon is at $r_{UH}\approx0.2$.
We can say that the mass of the conformal gravity BH decreases whereas the radius increases, on the other hand
the accretion rate is an increasing function of radius.
In this case, the critical points are overlapping at the universal horizon.
This implies that the mass of Schwarzschild de-Sitter BH decreases whereas the radius increases, on the other hand
the accretion rate is an increasing function of radius in the presence of cosmological constant $k$.

Now, the mass of Schwarzschild BH occurs in the bottom panel where we observe the maximum
accretion rate acquire for smaller radius. The critical points are overlapping at the
universal horizon but these points are away from the killing horizon in the
presence of mass function $\beta$. Four key points are observed for the Schwarzschild BH:
\begin{itemize}
  \item $\beta=0.7$, corresponding to $r_{UH}\approx 0.135$, $r_{KH}\approx 0.65$.
  \item $\beta=0.8$, corresponding to $r_{UH}\approx 0.133$, $r_{KH}\approx 0.70$.
  \item $\beta=0.9$, corresponding to $r_{UH}\approx 0.131$, $r_{KH}\approx 0.75$.
  \item $\beta=1.0$, corresponding to $r_{UH}\approx 0.129$, $r_{KH}\approx 0.80$.
\end{itemize}
The accretion rate of Schwarzschild BH is increasing
with the decreasing values of radius. So, it is a decreasing function of $r$.
Hence, we conclude that the mass of Schwarzschild BH is larger as compare
to conformal gravity BH and Schwarzschild de-Sitter BH.

\begin{figure}
\begin{center}
\includegraphics[width=80mm]{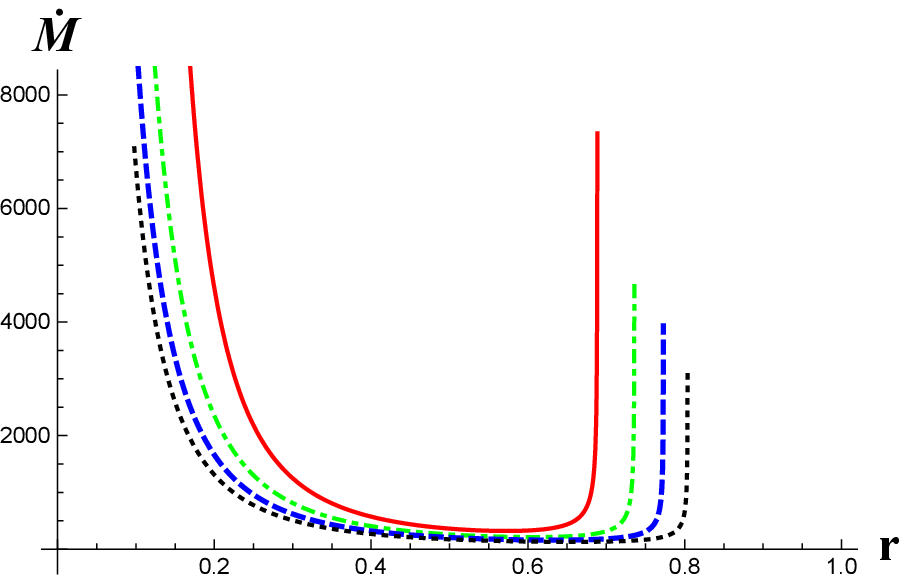}
\includegraphics[width=80mm]{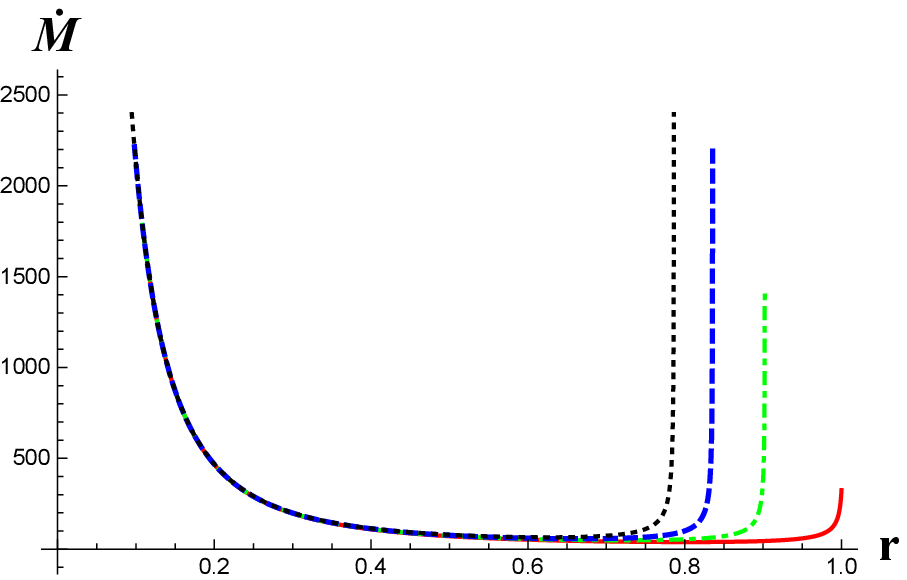}
\includegraphics[width=80mm]{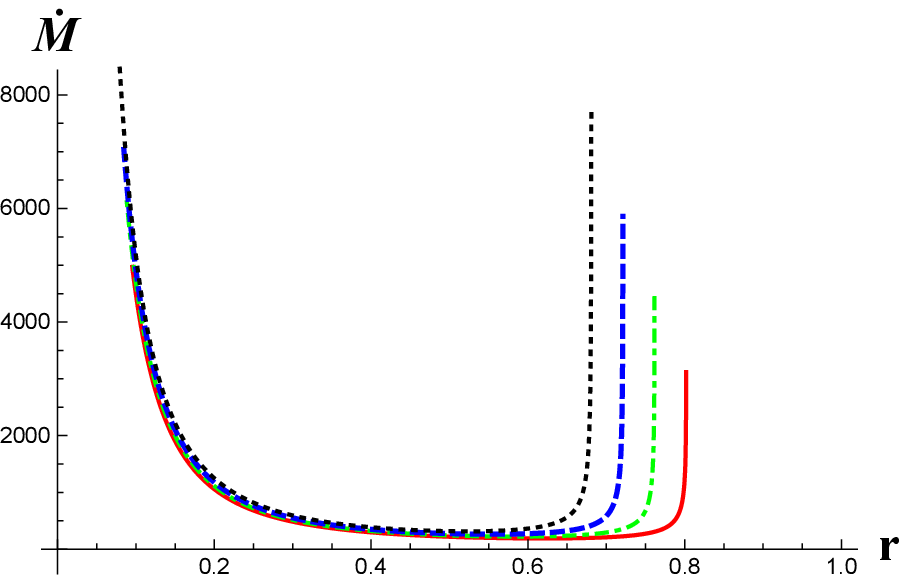}
\caption{In the mass accretion rate, left panel
(conformal gravity BH) displays the behavior of Eq. (\ref{84}) with
conformal parameters $\gamma=0.1$, $k=-1$, $\beta=1$. Right panel
(Schwarzschild de-Sitter BH) displays the behavior of Eq. (\ref{85})
with Schwarzschild de-Sitter BH parameters $\gamma=0$, $k=-1$, $\beta=1$.
Bottom panel (Schwarzschild BH) displays the behavior of Eq. (\ref{86}) with Schwarzschild BH
parameters $\gamma=0$, $k=0$, $\beta=1$. Other constants are taken as $L_0=0.90$, $L_1=0.5$.}
\end{center}
\end{figure}
\underline{\textbf{Exact solution for radiation fluid $\omega=1/3$:}}\\
By assuming $\omega=1$ in Eqs. (\ref{70}) and (\ref{74}), we calculate the radial velocity and the
energy-density of the ultra-stiff fluids, that is
\begin{eqnarray}\label{87}
u=&&\Big[\frac{\Big(-32A(r)L_1^4+\sqrt{1024A(r)^2L_1^8-27r^4L_0^6}\Big)^{1/3}}{4L_1^2}
\\\nonumber&&+\frac{3r^{4/3}L_0^2}{4L_1^{2/3}\Big(-32A(r)L_1^4+\sqrt{1024A(r)^2L_1^8-27r^4L_0^6}\Big)^{1/3}}\Big]^{2/3}.
\end{eqnarray}
The energy density of the fluid is given by
\begin{eqnarray}\label{88}
\varrho=&&\Big[\frac{L_1}{r^2}\Big]^{\frac{4}{3}}\Big[\frac{\Big(-32A(r)L_1^4+\sqrt{1024A(r)^2L_1^8-27r^4L_0^6}\Big)^{1/3}}{4L_1^2}
\\\nonumber&&+\frac{3r^{4/3}L_0^2}{4L_1^{2/3}\Big(-32A(r)L_1^4+\sqrt{1024A(r)^2L_1^8-27r^4L_0^6}\Big)^{1/3}}\Big]^{\frac{-8}{9}}.
\end{eqnarray}
The general form of the mass of BH is given by
\begin{eqnarray}\label{89}
\dot{M}=&&\Big[\frac{8\pi L_1^{\frac{7}{3}}}{r^{\frac{8}{3}}}\Big]\Big[\Big(-32A(r)L_1^4+\sqrt{1024A(r)^2L_1^8-27r^4L_0^6}\Big)^{1/3}\Big({4L_1^2}\Big)^{-1}+\frac{3r^{4/3}L_0^2}{4L_1^{2/3}}
\\\nonumber&&\times\Big(-32A(r)L_1^4+\sqrt{1024A(r)^2L_1^8-27r^4L_0^6}\Big)^{-1/3}\Big]^{\frac{-8}{9}}.
\end{eqnarray}
 1. The mass accretion rate of conformal gravity BH is
\begin{eqnarray}\label{90}
\dot{M}=&&\Big[\frac{8\pi L_1^{\frac{7}{3}}}{r^{\frac{8}{3}}}\Big]\Big[\Big(-32(1-\frac{\beta(2-3\beta\gamma)}{r}-3\beta\gamma+\gamma r-kr^2)L_1^4
\\\nonumber&&+\sqrt{1024\Big(1-\frac{\beta(2-3\beta\gamma)}{r}-3\beta\gamma+\gamma r-kr^2\Big)^2L_1^8-27r^4L_0^6}\Big)^{1/3}\Big({4L_1^2}\Big)^{-1}
\\\nonumber&&+\frac{3r^{4/3}L_0^2}{4L_1^{2/3}}\times\Big(-32(1-\frac{\beta(2-3\beta\gamma)}{r}-3\beta\gamma+\gamma r-kr^2)L_1^4 \\\nonumber&&+\sqrt{1024\Big(1-\frac{\beta(2-3\beta\gamma)}{r}-3\beta\gamma+\gamma r-kr^2\Big)^2L_1^8-27r^4L_0^6}\Big)^{-1/3}\Big]^{\frac{-8}{9}}.
\end{eqnarray}
 2. The mass accretion rate of Schwarzschild de-Sitter BH is

\begin{eqnarray}\label{91}
\dot{M}&&=\Big[\frac{8\pi L_1^{\frac{7}{3}}}{r^{\frac{8}{3}}}\Big]\Big[\Big(-32(1-\frac{2\beta}{r}-kr^2)L_1^4
\\\nonumber&&+\sqrt{1024\Big(1-\frac{2\beta}{r}-kr^2\Big)^2L_1^8-27r^4L_0^6}\Big)^{1/3}\Big({4L_1^2}\Big)^{-1}
\\\nonumber&&+\frac{3r^{4/3}L_0^2}{4L_1^{2/3}}\times\Big(-32(1-\frac{2\beta}{r}-kr^2))L_1^4
\\\nonumber&&+\sqrt{1024\Big(1-\frac{2\beta}{r}-kr^2\Big)^2L_1^8-27r^4L_0^6}\Big)^{-1/3}\Big]^{\frac{-8}{9}}.
\end{eqnarray}

 3. The mass accretion rate of Schwarzschild BH is

\begin{eqnarray}\label{92}
\dot{M}&&=\Big[\frac{8\pi L_1^{\frac{7}{3}}}{r^{\frac{8}{3}}}\Big]\Big[\Big(-32(1-\frac{2\beta}{r})L_1^4
\\\nonumber&&+\sqrt{1024\Big(1-\frac{2\beta}{r}\Big)^2L_1^8-27r^4L_0^6}\Big)^{1/3}\Big({4L_1^2}\Big)^{-1}
\\\nonumber&&+\frac{3r^{4/3}L_0^2}{4L_1^{2/3}}\times\Big(-32(1-\frac{2\beta}{r})L_1^4
\\\nonumber&&+\sqrt{1024\Big(1-\frac{2\beta}{r}\Big)^2L_1^8-27r^4L_0^6}\Big)^{-1/3}\Big]^{\frac{-8}{9}}.
\end{eqnarray}

In Fig. \textbf{10}, we plot mass accretion-rate $\dot{M}$ versus the radius $r$
for aforemention BHs for radiation fluid with the parameters $\gamma$, $k$
and $\beta$. The left panel (conformal gravity BH) shows that the accretion rate is
decreasing for smaller values of $r$, that is the red curve shows the minimum accretion rate
between the universal and Killing horizon. It is noted that the maximum accretion rate occurs
at the universal, but far from the  Killing horizon. It is the increasing function of the radius in
the presence of $\gamma$. The red curve in the right panel
depicts the minimum accretion rate at the universal and the Killing horizon for Schwarzschild
de-Sitter BH in the presence of cosmological constant $k$. It is noted here that the accretion
rate increases for smaller values of $k$ and we can see the maximum accretion rate near
$r\approx1.08, 0.82, 0.75, 0.69$ for $k=-1.0, -1.5, -2.0, -2.5$, respectively. It is the
decreasing function of radius that is mass increases when radius decreases.

We note that for Schwarzschild BH, the range of the maximum accretion rate is between the radius $0.5$ to $0.6$,
for larger value of $\beta=4$. It is also the decreasing function of the radius that is
accretion rate increases whereas the radius decreases.
So, four key points are observed for the Schwarzschild BH:
\emph{\begin{itemize}
  \item $\beta=1.0$, corresponding to $r_{UH}\approx 0.4$, $r_{KH}\approx 1.0$.
  \item $\beta=2.0$, corresponding to $r_{UH}\approx 0.4$, $r_{KH}\approx 0.90$.
  \item $\beta=3.0$, corresponding to $r_{UH}\approx 0.4$, $r_{KH}\approx 0.70$.
  \item $\beta=4.0$, corresponding to $r_{UH}\approx 0.4$, $r_{KH}\approx 0.60$.
\end{itemize}}
The mass accretion rate of Schwarzschild BH is increasing
and the radius decreases for increasing values of the mass function. Hence, we conclude that the mass of Schwarzschild BH is larger as compared to
conformal gravity BH and Schwarzschild de-Sitter BH.
\begin{figure}
\begin{center}
\includegraphics[width=80mm]{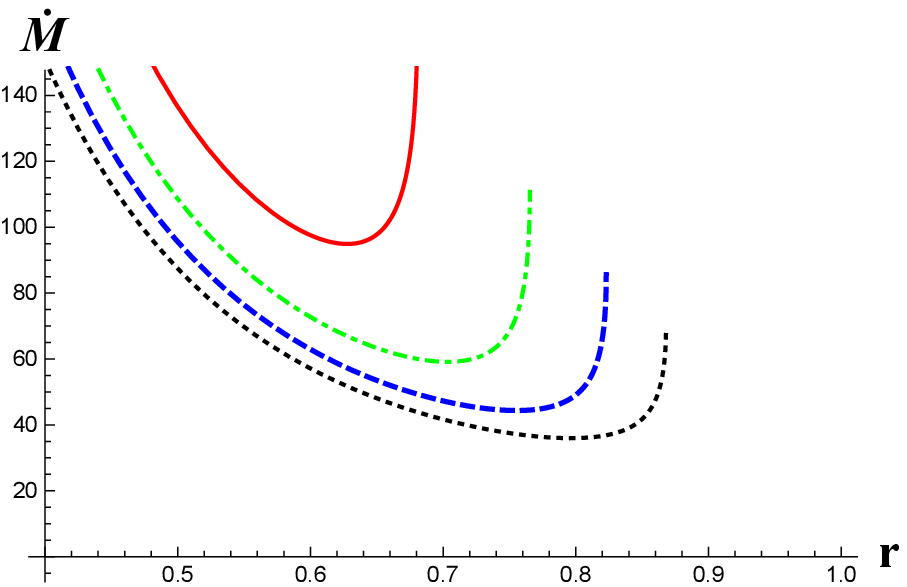}
\includegraphics[width=80mm]{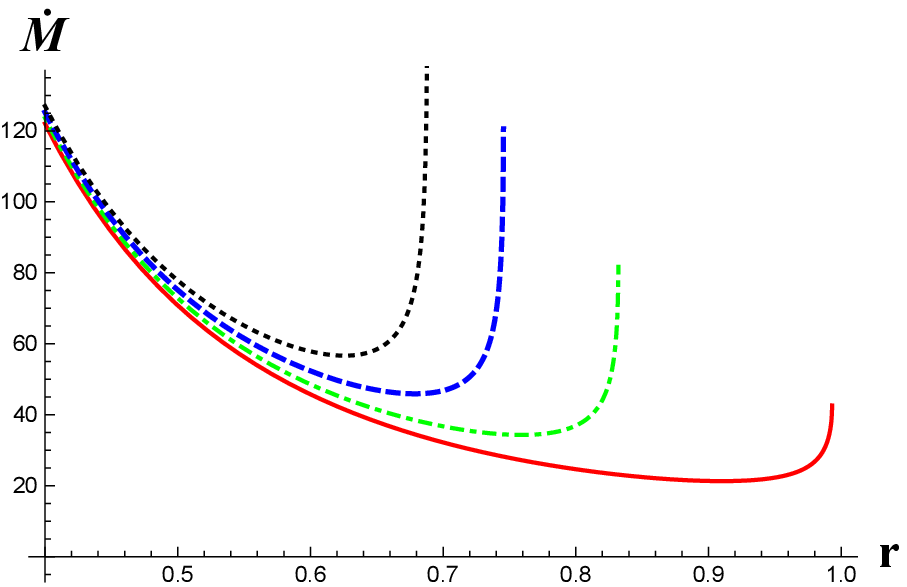}
\includegraphics[width=80mm]{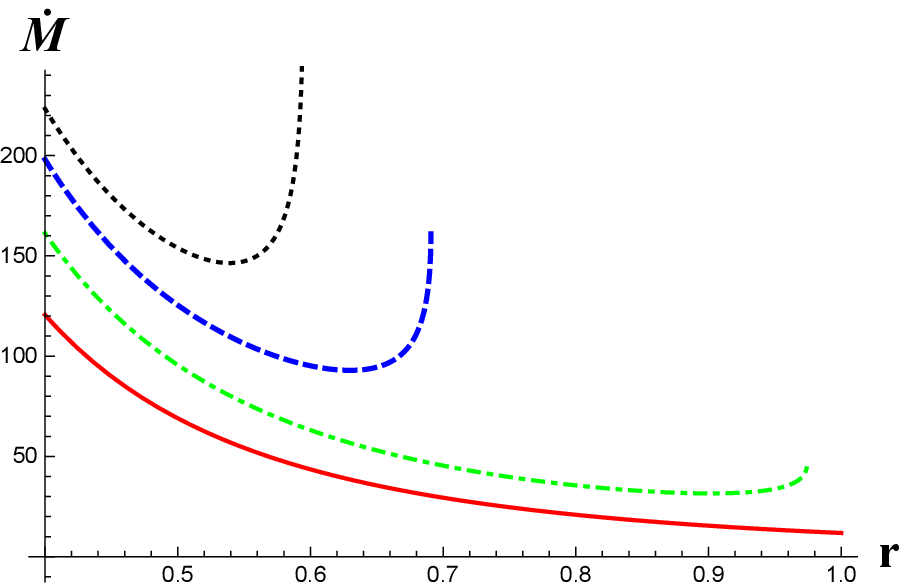}
\caption{For the mass accretion rate, left panel
(conformal gravity BH) displays the behavior of Eq. (\ref{89}) with
conformal parameters $\gamma=0.1$, $k=-1$, $\beta=1$. Right panel
(Schwarzschild de-Sitter BH) displays the behavior of Eq. (\ref{90})
with parameters $\gamma=0$, $k=-1$, $\beta=1$.
Bottom panel (Schwarzschild BH) displays the behavior of (\ref{91}) with the
parameters $\gamma=0$, $k=0$, $\beta=1$. Other constants are taken as $L_0=0.90$, $L_1=0.5$ .}
\end{center}
\end{figure}

\begin{figure}
\begin{center}
\includegraphics[width=80mm]{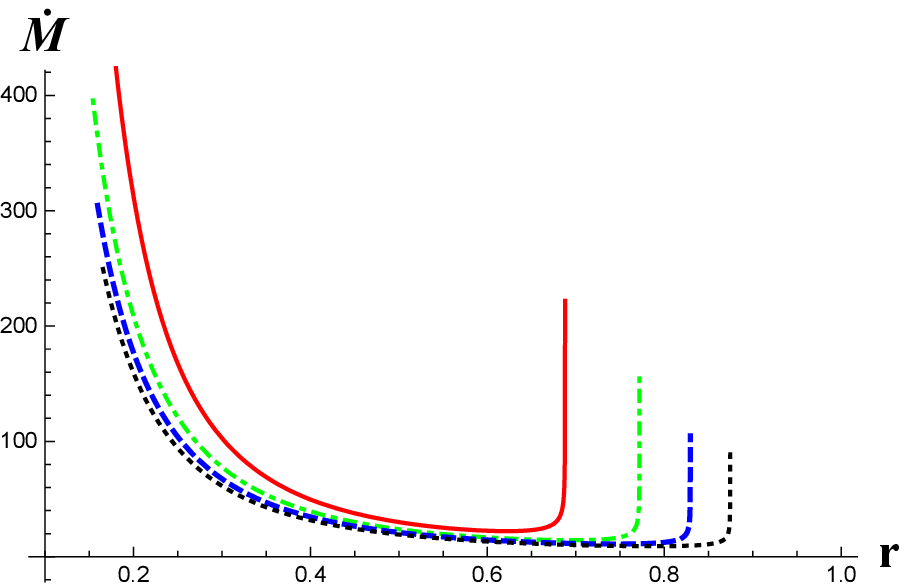}
\includegraphics[width=80mm]{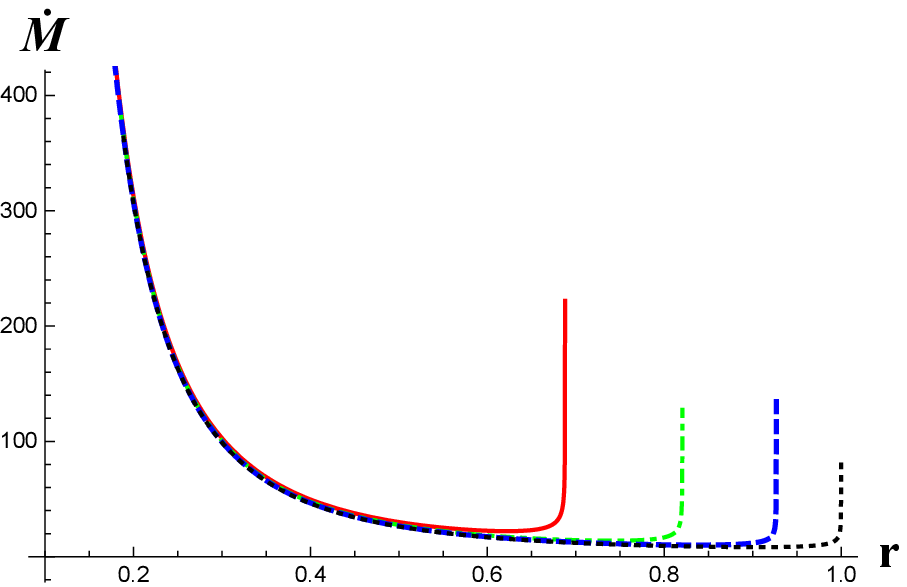}
\includegraphics[width=80mm]{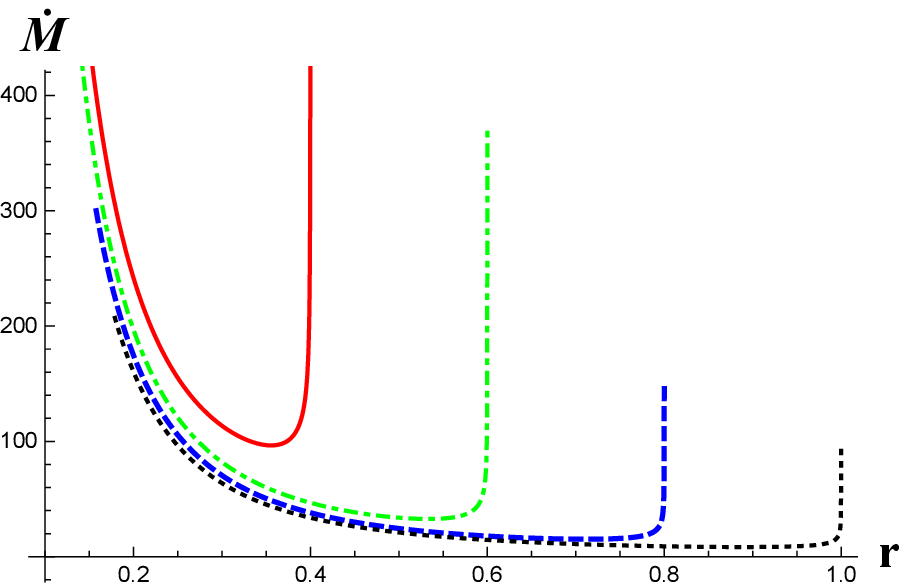}
\caption{For the mass accretion rate,
left panel displays the behavior of conformal gravity BH with
conformal parameters $\gamma=0.1$, $k=-1$, $\beta=1$.
Right panel displays the behavior of Schwarzschild de-Sitter BH
with parameters $\gamma=0$, $k=-1$, $\beta=1$. Bottom panel displays
the behavior of Schwarzschild BH with parameters
$\gamma=0$, $k=0$, $\beta=1$. Other constants are taken as $L_0=0.90$, $L_1=0.5$.}
\end{center}
\end{figure}
We plot mass accretion-rate ($\dot{M}$) versus the radius ($r$) for aforemention BHs
in the case of sub-relativistic fluid with particular values of conformal parameters $\gamma$, $k$
and $\beta$, as shown in Fig. \textbf{11}. The left panel (conformal gravity BH) shows that the accretion rate is
decreasing for larger values of $\gamma$. One can see the maximum accretion rate at $r\approx0.87$
for $\gamma=0.4$, $r\approx0.82$ for $\gamma=0.3$, $r\approx0.76$ for $\gamma=0.2$, $r\approx0.69$
for $\gamma=0.1$. While, the mass decreases but the radius increases in this case
between the universal and Killing horizon. The red curve in the right panel
depicts the maximum accretion rate between the universal and Killing horizon for Schwarzschild
de-Sitter BH in the presence of cosmological constant $k$. It is noted here that the accretion
rate decreases for smaller values of $k$ and we can see the maximum accretion rate near
$r\approx0.69, 0.82, 0.92, 1.0$ for $k=-1.0, -1.5, -2.0, -2.5$, respectively.
Now, the mass of Schwarzschild BH is given in the bottom panel where we observe the maximum
accretion rate for smaller radius between the universal and Killing horizons (see red curve).
The maximum accretion rate occurs between $r\approx0.2$ to $r\approx0.4$.
While, the minimum accretion rate occurs at the Killing horizon in the bottom panel.
We have noted that the minimum accretion rate is between $r\approx0.2$ to $r\approx1.0$
for the larger value of $\beta=4.0$. Four key points are observed for the Schwarzschild BH:
\begin{itemize}
  \item $\beta=1.0$, corresponding to $r_{UH}\approx 0.22$, $r_{KH}\approx 0.40$.
  \item $\beta=2.0$, corresponding to $r_{UH}\approx 0.21$, $r_{KH}\approx 0.60$.
  \item $\beta=3.0$, corresponding to $r_{UH}\approx 0.19$, $r_{KH}\approx 0.80$.
  \item $\beta=4.0$, corresponding to $r_{UH}\approx 0.18$, $r_{KH}\approx 1.0$.
\end{itemize}
The mass accretion rate of Schwarzschild BH decreases
with the increases of radius for the various values of the mass function.

\section{Conclusion}
%The world around us is actually not conformally invariant. For instance, in a conformally invariant
%theory, we cannot assess time intervals and lengths which is clearly not the case. If we desire
%to discover the possibility that conformally invariance is a basic system in Nature, we surely
%acknowledge that around us such a system is busted. When conformal invariance is automatically
%busted, Nature has appointed one of the likely vacua. The issue of spacetime singularities can be
%the explored inference that Nature can merely appoint a physical vacuum in the class of non-singularity
%metrics \cite{1m, 7w, 65, 66, 67}.

In this paper, we have investigated the spherically symmetric accretion around the conformal gravity BH,
 with four kinds of fluid as ultra-stiff fluid,
ultra-relativistic fluid, radiation-fluid and sub-relativistic fluid
by using the Hamiltonian approach. It is demonstrated that the energy density is always equal to pressure
in ultra-stiff fluids. In this case, it has been observed that supersonic as well as
subsonic accretion flow would exist for the particular values of the parameters. The critical radius in Schwarzschild BH
is larger than the conformal gravity BH and Schwarzschild de-Sitter BH.
The energy density is double of the pressure for ultra-relativistic fluid and there exists a
supersonic flow, which is followed by subsonic flow. The fluid flow around Schwarzschild BH
for ultra-relativistic fluid is entirely different as compared to
conformal gravity BH and Schwarzschild de-Sitter BH.
The $3D$-speed $v$ is very small but the radial distance is larger for Schwarzschild BH
as compare to conformal gravity BH and Schwarzschild de-Sitter BH.
It is also noted that the critical radius is very close
to the horizon for Schwarzschild BH as compare to conformal gravity BH and
Schwarzschild de-Sitter BH for ultra-relativistic fluid
(see Fig.\textbf{(4)}).
The nature of radiation-fluid and sub-relativistic fluid in which the energy density is greater than
the pressure is similar for $v>v_c$. A very simple behavior has been observed for the radiation-fluid that is
only supersonic flow exists for Schwarzschild BH while subsonic accretion exists for
conformal gravity BH and Schwarzschild de-Sitter BH. Further, for sub-relativistic fluid, the flow around Schwarzschild BH
is absolutely closer to ultra-relativistic fluid that is the critical radius is closer to the horizon
as compared to conformal gravity BH and Schwarzschild de-Sitter BH. The $3D$ speed for radial motion is
very small in Schwarzschild BH as compared to conformal gravity BH and Schwarzschild de-Sitter BH.

In addition, we have explored the results of mass accretion rate $\dot{M}$,
radial velocity $u$ and the energy density $\varrho$ corresponding Schwarzschild BH, conformal gravity BH and Schwarzschild de-Sitter BH.
We have investigated the mass accretion rate with
four types of fluid around conformal gravity BH, Schwarzschild de-Sitter BH and Schwarzschild BH
which is shown in Figs.\textbf{(8-11)}. The Schwarzschild BH
acquires the higher accretion rate as compare to conformal gravity BH and Schwarzschild de-Sitter BH
for ultra-stiff fluids. The mass accretion rate is smaller in Schwarzschild de-Sitter BH
than the conformal gravity BH and Schwarzschild BH for the ultra-relativistic fluid, radiation-fluid
and sub-relativistic fluid. It is concluded that the maximum mass accretion rate occur for conformal gravity BH at
$\dot{M}>8000$ versus $(r\leq0.8)$ for ultra-relativistic fluid.
The maximum mass accretion rate for Schwarzschild de-Sitter BH mass occurs at $\dot{M}>2500$ versus $(r\leq1.0)$
for ultra-relativistic fluid. Similarly, the maximum mass accretion rate for Schwarzschild BH occurs at $\dot{M}>8000$ versus $(r\leq0.8)$ for ultra-relativistic fluid.

Further, we have discussed the ultra-stiff,
ultra-relativistic, radiation and the sub-relativistic fluids with the equation of state which helps to identify that what kind of fluids is
accreting onto the BH. Moreover, critical points and conserved
quantities have been found for these fluids. The behavior of
accreting fluid has been discussed as subsonic and supersonic
according to equation of state. We have compared the fluid
flow for all models and observed that the fluid flow and CPs are
closer to Case $3$ instead of Case $2$ and $1$, respectively.

It has been analyzed that the subsonic accretion is followed
by the supersonic accretion inside the BH horizon and it does not support to the claim that "the flow must be supersonic at the
horizon" \cite{58}. So, for conformal gravity BH the fluid flow is neither supersonic
nor transonic near the horizon \cite{59,60} .
This outflow is unsteady because it follows a
subsonic path after passing through the saddle point ($r_c, v_c$) and
becomes supersonic with speed approaches to the speed of the light.
In the cosmological point of view, the point ($v=0, r=r_h$)
can be observed as repeller as well as  attractor
where the solution curves diverge and converge, respectively,
\cite{58x,64}. These results open a new window correspond to
fine-tuning and variability problems in dynamical systems.
\section*{Acknowledgments}
We are very grateful to the honorable referee, who put his/her efforts and give valuable suggestions for improving the manuscript.


\begin{thebibliography}{99}
\bibitem{1} Will, C. M.: Living Reviews in Relativity {\bf17} (2014) 4.
\bibitem{1e} Hayward, S. A.: Physical Review Letters {\bf96} (2006) 031103.
\bibitem{1f} Fan, Z. Y. and Wang, X.: Phys. Rev. D {\bf94} (2016) 124027.
\bibitem{1g} Toshmatov, B., Ahmedov, B., Abdujabbarov, A. and Stuchilk, Z.: Phys. Rev. D {\bf89} (2014) 104017.
\bibitem{1h} Mannheim, P. D.: Foundations of Physics {\bf42} (2012) 388.
\bibitem{1i} Bars, I., Steinhardt, P. and Turok, N.: Phys. Rev. D {\bf89} (2014) 043515.
\bibitem{1j} Bambi, C. and Modesto, L.: Phys. Lett. B {\bf721} (2013) 329.
\bibitem{1k} Bambi, C., Malafarina, D. and Modesto, L.: Phys. Rev. D {\bf88} (2013) 044009.
\bibitem{1l} Bambi, C., Malafarina, D. and Modesto, L.: Eur. Phys. J. C {\bf74} (2014) 2767.
\bibitem{1m} Bambi, C., Modesto, L. and Rachwal, L.: JCAP {\bf1709} (2017) 033.
\bibitem{1n} Bambi, C., Malafarina, D. and Modesto, L.: J. High Energy Physics {\bf2016} (2016) 147.
\bibitem{1o} Horava, P.: Phys. Rev. D {\bf79} (2009) 084008.
\bibitem{1p} Horava, P.: Physical Review Letters {\bf102} (2009) 161301 .
\bibitem{1pv} Maldacena, J.: arXiv:1105.5632.
\bibitem{1pw} Anastasiou, G. and Olea, R.: Phys. Rev. D {\bf94} (2016) 086008.
\bibitem{1px} Mannheim, P. D.: Gen. Relativ. Gravit. {\bf43} (2011) 703.
\bibitem{1py} Stelle, K. S.: Phys. Rev. D {\bf16} (1977) 953.
\bibitem{1py1} Mannheim, P. D. and Kazanas, D.: Astrophys. J {\bf342} (1989) 635.
\bibitem{1py2} Mannheim, P. D. and Kazanas, D.: Phys. Rev. D {\bf44} (1991) 417.
\bibitem{1py3} Mannheim, P. D.: Prog. Part. Nucl. Phys. {\bf56} (2006) 340.
\bibitem{1py4} Mannheim, P. D and O’Brien, J. G.: Phys. Rev. Lett.  {\bf106} (2011) 121101.
\bibitem{1pz1} Lu, H., Pang, Y., Pope, C. N. and Vazquez-Poritz, J. F.: Phys. Rev. D {\bf86} (2012) 044011.
\bibitem{1pz2} Xu, W. and Zhao, L.: Phys. Lett. B {\bf736} (2014) 214.
\bibitem{1pz3} Xu, H., Sun, Y. and Zhao, L.: Int. J. Mod. Phys. D {\bf13} (2017) 1750151.
\bibitem{1pz4} Xu, H. and Yung, M. H.: Phys. Lett. B {\bf783} (2018) 36.
\bibitem{2} Hioki, K and Maeda, K. I.: Phys. Rev. D {\bf80} (2009) 024042.
\bibitem{2a} Hagihara, Y. and Japan, J.:  Astron. Geophys. {\bf8} (1931) 67.
\bibitem{2b} Chandrasekhar, S.: The Mathematical Theory of Black Holes (Oxford University Press, Oxford,
1983).
\bibitem{2c} Hackmann, E. and Lammerzahl, C.: Phys. Rev. Lett. {\bf100} (2008) 171101.
\bibitem{2d} Hackmann, E., Kagramanova, V., Kunz, J. and Lammerzahl, C.: Phys. Rev. D {\bf78} (2008) 024035.
\bibitem{2e} Hackmann, E., Kagramanova, V., Kunz, J. and Lammerzahl, C.: Phys. Rev. D {\bf78} (2008) 124018
\bibitem{2f} Grunau, S., Kagramanova, V., Kunz, J. and Lammerzahl, C.: Phys. Rev. D {\bf86} (2012) 104002.
%\bibitem{2g} Soroushfar, S., Saffari, R., Kunz, J. and Lammerzahl, C.: Phys. Rev. D {\bf92} (2015) 044010.
%\bibitem{2h} Soroushfar, S., Saffari, R. and Sahami, E.: [arXiv:1601.03143 [gr-qc]].
%\bibitem{2i} Soroushfar, S., Saffari, R. and Jafari, A.: [arXiv:1512.08449 [gr-qc]].
%\bibitem{2j} Hoseini, B. et al.: arXiv:1602.03898 [gr-qc].
%\bibitem{2k} Soroushfar, S. et al.: arXiv:1605.08976 [gr-qc].
%\bibitem{2l} Soroushfar, S., Saffari, R. and Grunau, S.: arXiv:1605.08975 [gr-qc].
%\bibitem{2m} Kazempour, S., Saffari, R. and Soroushfar, S.: arXiv:1606.06106 [gr-qc].
\bibitem{3} Barabash, O. V. and Shtanov, V. Y.: Phys. Rev. D {\bf60} (1999) 064008.
\bibitem{4} Brihaye, Y. and Verbin, Y.: Phys. Rev. D {\bf80} (2009) 124048.
\bibitem{5} Wood, J. and Moreau, W.: gr-qc/0102056.
\bibitem{6} Edery, A. Methot, A. A. and Paranjape, M. B.: Gen. Rel. Grav. {\bf33} (2001) 2075.
\bibitem{7} Sultana, J. Kazanas, D. and Said, J. L.: Phys. Rev. D {\bf86} (2012) 084008.
\bibitem{7u} Toshmatov, B., Ahmedov, B., Abdujabbarov, A. and Bambi, C.: Phys. Rev. D {\bf97} (2018) 124005.
\bibitem{7v} Zhou, M., et al.:  Phys. Rev. D {\bf98} (2018) 024007.
\bibitem{7w} Toshmatov, B., Bambi, C. Ahmedov, B., Abdujabbarov, A. and Stuchilk, Z.: Eur. Phys. J. C {\bf77} (2017) 542.
\bibitem{7x} Toshmatov, B., Bambi, C. Ahmedov, B., Abdujabbarov, A., Stuchilk, Z. and Schee, J.: Phys. Rev. D {\bf96} (2017) 064028.
\bibitem{7y} Haydarov, K., Abdujabbarov, A., Rayimbaev, J. and Ahmedov, B.: Universe {\bf6} (2020) 6030044.
\bibitem{8} Jamil, M. Rashid, M. A. and Qadir, A.: Eur. Phys. J. C {\bf58} (2008) 325.
\bibitem{9} Jamil, M.: Eur. Phys. J. C {\bf 62} (2009) 609.
\bibitem{10} Hoyle, F. and Lyttelton, R. A.: Proc. Camb. Philos. Soc. {\bf35} (1939) 405.
\bibitem{11} Bondi, H. and Hoyle, F.: Mon. Not. R. Astron. Soc. {\bf104} (1944) 273.
\bibitem{12} Bondi, H.: Mon. Not. Roy. Astron. Soc. {\bf112} (1952) 195.

\bibitem{15} Michel, C. F.: Astrophys. Space Sci. {\bf15} (1972) 153
\bibitem{13} Shapiro, S. L. and Teukolsky, S. A.: Black Holes, White Dwarfs and Neutron Stars Wiley, New
             York, (1983) pp 305.
\bibitem{14} Frank, J., King, A. and King, D.: Accretion Power in Astrophysics (Cambridge University
             Press, Cambridge, 2002).
\bibitem{16} Begelman, C. M.: Astron. Astrophys. {\bf70} (1978)583.
\bibitem{17} Bettwieser, E. and Glatzel, W.: Astron. Astrophys. {\bf94} (1981)306.
\bibitem{18} Thorne,  K. S. Flammang, R. A. and Zytkow, A. N.: Mon. Not. R. Acad. Sci. {\bf194} (1981)475.
\bibitem{19} Pandey, S. U.: Astrophys. Space Science {\bf136} (1987) 195.
\bibitem{20} Harko, T. and  Mak, K. M.: Phys. Rev. D {\bf319} (2005) 471.
\bibitem{21} Harko, T. and  Mak, K. M.: Phys. Rev. D {\bf636} (2006) 8.
\bibitem{22} Perlmutter, S. et al.: Astrophys. J {\bf 517} (1999) 565.
\bibitem{24} Babichev, E. O. Dokuchaev, V. I. and  Eroshenko, Yu. N.: Phys. Usp. {\bf100} (2013) 1155.
\bibitem{25} Debnath, U.: Eur. Phys. J. C {\bf75} (2015) 129.
\bibitem{26} Babichev, E., Dokuchaev, V. and Eroshenko, Y.: Phys. Rev. Lett {\bf93} (2004) 021102.
\bibitem{27} Babichev, E., Dokuchaev, V. and Eroshenko, Y.: J. Exp. Theor. Phys. {\bf100} (2005)528.
\bibitem{28} Gao, G., Chen, X., Faraoni, V. and Shen, Y. G.: Phys. Rev. D {\bf78} (2008) 024008.
\bibitem{29} John, A. J., Ghosh, S. G. and Maharaj, S. D.: Phys. Rev. D {\b88} (2013) 104005.
\bibitem{30} Ganguly, A., Ghosh, S. G. and Maharaj, S. D.: Phys. Rev. D {\b90} (2014) 064037.
\bibitem{31} Karkowski, J. and Malec, E.: Phys. Rev. {\bf87} (2013) 044007.
\bibitem{32} Mach, P. and Malec, E.: Phys. Rev. {\bf88} (2013) 084055.
\bibitem{33} Guzman, F. S. and Lora-Clavijo, F. D.: MNRAS. {\bf225} (2011) 415.
\bibitem{34} Ananda, D. B. Bhattacharya, S. and Das, T. K.: Gen. Relat. Gravit. {\bf47} (2015) 96.
\bibitem{35} Sharif, M. and Iftikhar, S.: Eur. Phys. J. C {\bf76} (2016) 147.
\bibitem{36} Sharif, M. and Iftikhar, S.: Eur. Phys. J. C {\bf76} (2016) 404.
\bibitem{37} Sharif, M. and Iftikhar, S.: Eur. Phys. J. C {\bf76} (2016) 630.
\bibitem{38} Sharif, M. and Shahzadi, M.: Eur. Phys. J. C {\bf77} (2017) 363.
\bibitem{39} Sharif, M. and Mumtaz, S. Not. R.: Astro. Soc. {\bf471} (2017) 1215.
\bibitem{40} Sharif, M. and Abbas, G.: Mod. Phys. Lett. A {\bf26} (2011) 1731.
\bibitem{41} Sharif, M. and Abbas, G.: Chin. Phys. Lett. A {\bf29} (2012) 010401.
\bibitem{41a} Chaverra, E. and Sarbach, O.: Class. Quant. Gravity {\bf32} (2015) 15.
\bibitem{41b} Yang, R.: Physical Review D {\bf92} (2015) 8.
\bibitem{41c} Gangopadhyay, S., Paik, B. and Mandal, R.: Int. J.
Mod. Phy. A {\bf33} (2018) 1850084.
\bibitem{41d} John, A.J.: Monthly Notices of the Royal Astronomical Society {\bf490} (2019) 3.
\bibitem{42} Ahmad, A. K. Azreg-Ainou, M. Faizal, M and Jamil, M.: Eur. Phys. J. C. {\bf76} (2016) 280.
\bibitem{43} Ahmad, A. K. Azreg-Ainou, M. Bahamonde, S. Capozziello, S and Jamil, M.: Eur. Phys. J. C {\bf76} (2016) 269.
\bibitem{44} Jawad, A. and Shahzad, M.U.: Eur. Phys. J. C {\bf77} (2017) 515.
\bibitem{45} Abbas, G. and Ditta, A.: Mod. Phys. Lett. A {\bf33} (2018) 1850070.
\bibitem{46} Abbas, G. and Ditta, A.: Gen. Relat. Gravit. {\bf51} (2019) 43.
\bibitem{47} Abbas, G., Ditta, A., Jawad, A. and Umair, S.: Gen. Relat. Gravit. {\bf51}(2019) 136.
\bibitem{48} Ditta, A. and Abbas, G.: Chinese Journal of Physics {\bf65} (2020) 333.
\bibitem{49} Umair, S., Ali, R., Jawad, A. and Rani, S.: Chinese Physics C {\bf44}(2020) 065106.
\bibitem{55} Rezzolla, L., Zanotti, O.: Relativistic Hydrodynamics, Oxford University
Press, London (2013).
\bibitem{56} Ficek, F.: Class. Quantum. Grav. {\bf32} (2015) 235008.
\bibitem{57} Bahamonde, S. and Jamil, M.: Eur. Phys. J. C {\bf75} (2015) 508.
\bibitem{58} Chakrabarti, S. K.: Int. J. of Mod. Phys. D \textbf{20}(2011) 1723.
\bibitem{58x} Ahmed, K. et al.: Eur. Phys. J. C \textbf{76} (2016) 280.
\bibitem{59} Novikov, I., Thorne,  K.S.: in Black Holes, ed. by C.
DeWitt, B. De Witt (Gordon and Breach, New York, 1973).
\bibitem{60} Chakrabarti, S.K.: Theory of transonic astrophysical flows (World
Scientific, Singapore, 1990).
\bibitem{61}Nagle, R. K., Saff, E. B., Snider, A. D.:
Fundamentals of differential equations and boundary value problems,
6th edn. (Pearson, International Edition, UK, 2012).
\bibitem{62}
Polking, J., Boggess, A., Arnold, D.: Diffrential equations with
boundary value problems, 2nd edn. (Prentice Hall, Upper Saddle
River, 2006).
\bibitem{63} Bugl, P.: Differential Equations: Matrices and Models (Prentice Hall,
Englewood Cliffs, 1995).
\bibitem{64}  Azreg-Ainou, M.: Class. Quantum
Gravity \textbf{30} (2013)205001.
%\bibitem{65} Chakrabarty, H., Benavides-Gallego, C. A., Bambi, C. and Modesto, L.: JHEP {\bf1803} (2018) 013.
%\bibitem{66} Bambi, C., Modesto, L., Porey, S. and Rachwal, L.: Eur. Phys. J. C {\bf78} (2018) 116.
%\bibitem{67} Zhang, Q., Modesto, L. and Bambi, C.: Eur. Phys. J. C {\bf78} (2018) 506.
\end{thebibliography}
\end{document}